\documentclass[10pt,letterpaper]{article}
\usepackage[top=0.85in,left=2.75in,footskip=0.75in,marginparwidth=2in]{geometry}

\usepackage[utf8]{inputenc}
\usepackage{cite}
\usepackage{nameref,hyperref}

\usepackage{microtype}
\DisableLigatures[f]{encoding = *, family = * }

\raggedright
\usepackage{changepage}
\usepackage[aboveskip=1pt,labelfont=bf,labelsep=period,singlelinecheck=off]{caption}

\makeatletter
\renewcommand{\@biblabel}[1]{\quad#1.}
\makeatother

\usepackage{lastpage,fancyhdr,graphicx}
\usepackage{epstopdf}
\fancyheadoffset[L]{2.25in}
\fancyfootoffset[L]{2.25in}

\usepackage{color}
\definecolor{Gray}{gray}{.25}

\usepackage{graphicx}
\usepackage{sidecap}
\usepackage{wrapfig}
\usepackage[pscoord]{eso-pic}
\usepackage[fulladjust]{marginnote}
\reversemarginpar

\usepackage{amsmath,amssymb,amsfonts,amstext,mathtools,bm,latexsym}
\usepackage{siunitx}
\usepackage{float}
\usepackage{booktabs}
\usepackage{multirow}
\usepackage{subcaption}
\usepackage{algorithm}
\usepackage{algpseudocode}
\usepackage{todonotes}

\newcommand{\ignore}[1]{}  
\DeclareMathOperator*{\argmin}{arg\,min}

\usepackage{tikz}
\usetikzlibrary{arrows.meta,positioning,calc}
\usepackage{tabularx}
\usepackage{optidef}
\usepackage{array}
\usepackage{amsthm}
\usepackage{cleveref}
\labelformat{subsection}{\thesection.#1}

\begin{document}
\vspace*{0.35in}

\begin{flushleft}
{\Large
\textbf\newline{Meta-Reinforcement Learning for Robust and Non-greedy Control Barrier Functions in Spacecraft Proximity Operations}
}
\newline
\\
Minduli C. Wijayatunga\textsuperscript{1,*},
Richard Linares\textsuperscript{2},
Roberto Armellin\textsuperscript{3}
\\
\bigskip
\bf{1} University of Illinois Urbana--Champaign, Urbana, IL, USA
\\
\bf{2} Massachusetts Institute of Technology, Cambridge, MA, USA
\\
\bf{3} University of Auckland, Auckland, New Zealand
\\
\bigskip
* corresponding: minduli@illinois.edu

\end{flushleft}

\section*{Abstract}
Autonomous spacecraft inspection and docking missions require controllers that guarantee safety under thrust constraints and uncertainty. Input-constrained control barrier functions (ICCBFs) provide a framework for safety certification under bounded actuation; however, conventional ICCBF formulations can be overly conservative and exhibit limited robustness to uncertainty, leading to high fuel consumption and reduced mission feasibility. This paper proposes a framework in which the full hierarchy of class-$\mathcal{K}$ functions defining the ICCBF recursion is parameterized and learned, enabling localized shaping of the safe set and reduced conservatism. A control margin is computed efficiently using differential algebra to enable the learned continuous-time ICCBFs to be implemented on time-sampled dynamical systems typical of spacecraft proximity operations. A meta-reinforcement learning scheme is developed to train a policy that generates ICCBF parameters over a distribution of hidden physical parameters and uncertainties, using both multilayer perceptron (MLP) and recurrent neural network (RNN) architectures. Simulation results on cruise control, spacecraft inspection, and docking scenarios demonstrate that the proposed approach maintains safety while reducing fuel consumption and improving feasibility relative to fixed class-$\mathcal{K}$ ICCBFs, with the RNN showing a particularly strong advantage in the more complex inspection case.

\section*{Keywords}
Control Barrier Functions,  Meta-Reinforcement Learning, Time-Sampled Systems, Spacecraft Proximity Operations


\section{Introduction}
Spacecraft rendezvous and proximity operations (RPO) must satisfy stringent safety constraints while operating under thrust and onboard-computation limits. They require controllers that can maintain safety without sacrificing fuel or time efficiency, or robustness to uncertainty, especially when operating autonomously. Achieving this balance remains challenging with conventional control architectures, which often separate planning and safety enforcement and provide limited guarantees under real flight execution constraints.

Control Barrier Functions (CBFs) provide a set-theoretic approach to safety certification by enforcing the forward invariance of a prescribed safe set, typically implemented via a real-time quadratic program (CBF–QP) that minimally modifies a nominal command \cite{XiaoCassandrasBelta2023SafeAutonomyCBF}. This structure is attractive for autonomous RPO as it provides safety guarantees and computational tractability. However, several characteristics of RPO limit the effectiveness of conventional CBF–QP safety filters.

(1) \emph{Relative-degree limitations.}
Many RPO constraints such as approach cones, Line-Of-Sight (LOS) constraints, Keep-Out-Zone (KOZ), and Keep-In-Zone (KIZ) are geometric and depend only on relative position. Under effectively second-order translational dynamics present in these missions, for such barriers, the control does not appear in the first time derivative of the barrier, so $L_g h \equiv 0$. Thus, a standard first-order CBF--QP cannot directly shape the trajectory evolution without additional constructions.

(2) \emph{Thrust constraints.}
Conventional CBF formulations do not account for thrust constraints. As a result, the nominal safe set defined by the barrier function may be unattainable when thrust limits are enforced, as there can exist states that satisfy the barrier constraint but for which no admissible control input can maintain forward invariance. 

(3) \emph{Time-sampled implementation.}
Although spacecraft dynamics evolve continuously in time, spacecraft thrust commands are typically executed as a time-sampled controller with a zero-order hold. Since classical CBF conditions are derived in continuous time, applying them directly to time-sampled dynamical systems can forfeit safety guarantees between samples unless inter-sample effects are taken into account \cite{9417092}.

(4) \emph{Myopic behaviour.}
Even when feasible, CBF--QP safety filters are typically \emph{myopic} as they enforce constraints instantaneously without anticipating future activations or optimising objectives beyond safety. This can yield oscillatory or jittery control \cite{zhang2023advancedsafetyfilter}, excessive thrust expenditure, and suboptimal trajectories where locally safe actions conflict with globally efficient strategies.

(5) \emph{Model uncertainty.}
RPO often involves substantial modeling uncertainty and partially unknown/hidden parameters, particularly in uncooperative or debris-interaction scenarios \cite{VolpeCirciSabatiniPalmerini2022GNCUncooperativeTumbling}. Since standard CBF conditions assume a known control-affine model, unmodelled dynamics and parametric uncertainty can invalidate barrier derivative bounds and compromise invariance unless robustness mechanisms are introduced—often at the expense of feasibility under tight actuation limits.

Fortunately, a growing body of recent work has sought to mitigate these shortcomings.  To handle the relative degree limitations, \emph{Higher-Order} CBF constructions differentiate the barrier condition until the control appears explicitly, enabling a QP-based filter to influence barrier evolution \cite{9456981}. Input-Constrained CBFs (ICCBFs) do this as well, and also incorporate thrust limits directly into the barrier construction by constructing an input-admissible inner safe set \cite{agrawal2021safe}. While these extensions improve applicability to spacecraft-relevant geometric constraints, they can introduce conservatism by shrinking the admissible set and tightening feasibility margins; near constraint boundaries or under disturbances, the resulting QP optimizations can still become infeasible \cite{10885940}.

To address the time-sampling problem, discrete-time CBF variants \cite{agrawal2017discrete,9304281,doi:10.2514/6.2024-4935,9483029} have been introduced to enforce safety at sampling times, but they often yield non-affine control dependence, with higher-order constraints further increasing computational complexity. To retain the tractability of a single convex QP per guidance cycle while using continuous-time CBF formulations in time-sampled systems, {margin-based} CBF implementations \cite{oruganti2023robustsampleddataCBF,9417092} have been used to strengthen the CBF inequality using explicit inter-sample bounds derived via Lipschitz/reachable-set arguments. However, for larger update periods, these margins can significantly shrink the admissible set, and their computation often requires numerically intensive minimization to obtain Lipschitz constants per time step.

Compared with other limitations, the literature addressing \emph{myopic} behavior in CBF formulations is relatively limited. In \cite{zeng2021enhancing,9483029}, myopia is mitigated by embedding CBF constraints within Model Predictive Control (MPC) formulations, and \cite{breeden2022predictive} utilizes predictive safety filtering for cost-awareness. These approaches can improve long-horizon behavior, but introduce additional design choices (e.g., horizon and weight selection) and often require more complex online optimization than a per-step CBF--QP, which can be challenging under flight-relevant update rates and onboard compute constraints.

To address the challenge of model and parameter uncertainty, uncertainty-aware CBF variants \cite{kolathaya2019input,xu2015robustness,dean2020perception} have been developed.  Complementary robust time-sampled approaches \cite{breeden2022robust,oruganti2024robust} have also been developed to incorporate inter-sample margins, thereby retaining safety under zero-order hold and bounded disturbances. Aerospace applications have begun adopting these ideas for uncertainty-aware constrained trajectory generation \cite{breeden2023robustsat}. However, these mechanisms can trade robustness for increased conservatism, which can reduce feasibility under tight actuation limits, motivating learning-augmented designs that adapt barrier parameters to ensure fuel and time-optimal trajectories. 

One such approach is presented in~\cite{11128840}, which introduces an uncertainty-aware online adaptation framework for discrete-time ICCBFs by combining probabilistic learning with MPC-based verification to refine class-$\mathcal{K}$ parameters. This framework provides strong safety guarantees through explicit online validation, but relies on multi-step prediction and optimization at each control update. In contrast, the present work considers continuous-time ICCBFs executed under time-sampled zero-order hold and targets low-overhead online adaptation suitable for spacecraft guidance rates. Specifically, ICCBFs are used to encode thrust constraints, while time-sampled execution is handled via a control safety margin computed efficiently using Differential Algebra (DA), avoiding the computationally intensive minimization-based margin estimation used in prior work. 

The key idea of this work is to parameterize and learn the full class-$\mathcal{K}$ hierarchy in the ICCBF recursion, exposing the degrees of freedom that govern conservatism and the input-admissible inner safe set. A recurrent meta-RL policy is trained over large distributions of hidden physical and operational parameters, along with state and thrust uncertainties, to produce task-adaptive class-$\mathcal{K}$ parameters online while preserving real-time tractability via a single convex QP per control step. The main contributions of this work are as follows.
\begin{enumerate}
    \item \textbf{Learnable ICCBF recursion:} Parameterize and learn the class-$\mathcal{K}$ hierarchy to reduce conservatism and increase feasibility.
    \item \textbf{DA-based time-sampled margin:} Incorporate an inter-sample safety margin into the ICCBF--QP and compute it efficiently using DA, preserving real-time tractability.
    \item \textbf{Meta-RL adaptation under uncertainty:} Train a meta-RL policy over large variations of hidden parameters and disturbance distributions to produce robust, transferable ICCBF parameters. 
\end{enumerate}

To validate the developed framework, Monte Carlo (MC) studies are conducted on three tasks of increasing dimensionality: a one-dimensional cruise-control problem, a two-dimensional docking problem, and a three-dimensional inspection problem. In all cases, the objective is to conserve fuel while maintaining safety and reaching the target; the inspection case additionally seeks to maximise an inspection metric. Meta-RL training is performed using both a standard MLP policy and a recurrent Long Short-Term Memory (LSTM) policy for comparison.

The remainder of the paper is organized as follows. Section~\ref{sec:prelims} reviews the problem setup, time-sampled dynamical models, and preliminaries on CBFs, ICCBFs, and meta-RL. Section~\ref{sec:method} presents the proposed methodology, including the parameterization of the ICCBF recursion, control margin computation, and the associated QP formulations. Section~\ref{sec:learning} describes the meta-RL training framework, including the recurrent policy architecture, reward design, and observation structure. Section~\ref{sec:experiments} reports results for the three case studies, and Section~\ref{sec:conclusion} concludes the paper.

\section{Theoretical Background} \label{sec:prelims}
\textbf{Notation.} $\mathbb{R}$ and $\mathbb{R}_+$ denote the real and non-negative real numbers. A function $\alpha:[0,a)\to[0,\infty)$ is class-$\mathcal{K}$ if it is continuous, strictly increasing, and $\alpha(0)=0$. Lie derivatives are denoted
$L_{\boldsymbol{f}} h(\boldsymbol{x})=\frac{\partial h}{\partial \boldsymbol{x}}\boldsymbol{f}(\boldsymbol{x})$ and $L_{\boldsymbol{g}} h(\boldsymbol{x})=\frac{\partial h}{\partial \boldsymbol{x}}\boldsymbol{g}(\boldsymbol{x})$.

\subsection{Problem Setup}
Consider the control-affine system
\begin{equation}\label{xdoteq}
\dot{\boldsymbol{x}}=\boldsymbol{f}(\boldsymbol{x})+\boldsymbol{g}(\boldsymbol{x})\boldsymbol{u},\qquad \boldsymbol{x}\in\mathcal{\boldsymbol{X}}\subset\mathbb{R}^n,\; \boldsymbol{u}\in\mathcal{U}\subset\mathbb{R}^m,
\end{equation}
where $\boldsymbol{f}$ and $\boldsymbol{g}$ are sufficiently smooth. Safety is specified by a continuously differentiable function $h:\mathcal{\boldsymbol{X}}\to\mathbb{R}$ through the safe set
\begin{equation}
\mathcal{S}\triangleq\{\boldsymbol{x}\in\mathcal{\boldsymbol{X}}: h(\boldsymbol{x})\ge 0\}.
\end{equation}
A feedback law renders $\mathcal{S}$ \emph{forward invariant} if $\boldsymbol{x}(0)\in\mathcal{S}$ implies $\boldsymbol{x}(t)\in\mathcal{S}$ for all $t\ge 0$.

\subsection{Time-sampled Execution and Zero-Order Hold}
Although \eqref{xdoteq} evolves in continuous time, sensing and control updates are typically executed at discrete instants on digital hardware. Let $t_k=kT$ denote sampling instants, where $T>0$ is the sampling time step. Under a time-sampled implementation with ZOH, the control input is updated only at $t_k$ and held constant between updates:
\begin{equation}
\boldsymbol{u}(t)=\boldsymbol{u}_k,\qquad t\in[t_k,t_{k+1}),\qquad \boldsymbol{u}_k\in\mathcal{U},
\label{eq:zoh}
\end{equation}
where $\boldsymbol{u}_k$ is computed from the sampled state $\boldsymbol{x}_k \triangleq \boldsymbol{x}(t_k)$. The resulting trajectory segment on $[t_k,t_{k+1})$ is the solution of \eqref{xdoteq} with initial condition $\boldsymbol{x}(t_k)=\boldsymbol{x}_k$ and constant input $\boldsymbol{u}_k$. The state update can be written in integral form as
\begin{equation}
\boldsymbol{x}_{k+1}=\boldsymbol{x}_k+\int_{0}^{T}\!\Big(\boldsymbol{f}\big(\boldsymbol{x}(t_k+\tau)\big)+\boldsymbol{g}\big(\boldsymbol{x}(t_k+\tau)\big)\boldsymbol{u}_k\Big)\,d\tau.
\label{eq:flow_integral}
\end{equation}
Note that in this setting, enforcing a continuous-time CBF only at the sampling instants does not guarantee safety on $(t_k,t_{k+1})$ unless inter-sample effects are explicitly bounded or incorporated into the condition \cite{breeden2023robustsat}.

\subsection{Control Barrier Functions}
A standard (relative-degree-one) CBF condition enforces forward invariance by requiring the existence of a class-$\mathcal{K}$ function $\alpha$ such that
\begin{equation}\label{eq:cbf}
\sup_{\boldsymbol{u}\in\mathcal{U}}\big[L_{\boldsymbol{f}} h(\boldsymbol{x})+L_{\boldsymbol{g}} h(\boldsymbol{x})\boldsymbol{u}\big]\ge -\alpha\!\left(h(\boldsymbol{x})\right),\qquad \forall \boldsymbol{x}\in\mathcal{X}.
\end{equation}
Equivalently, any control satisfying the pointwise inequality
\begin{equation}\label{eq:cbf-ineq}
L_{\boldsymbol{f}} h(\boldsymbol{x})+L_{\boldsymbol{g}} h(\boldsymbol{x})\boldsymbol{u}+\alpha\!\left(h(\boldsymbol{x})\right)\ge 0
\end{equation}
renders $\mathcal{S}$ forward invariant under standard regularity assumptions (see, e.g., \cite{agrawal2021safe,8796030}).

\subsection{Control Lyapunov Functions}
For some tasks presented in this work (cruise control and docking), goal-reaching is encoded via a Control Lyapunov Function (CLF) inequality alongside the CBF constraint in a QP. Let $V:\mathcal{X}\to\mathbb{R}_+$ be continuously differentiable. A standard CLF condition requires the existence of a class-$\mathcal{K}$ function $\alpha_V$ such that
\begin{equation}\label{eq:clf}
\inf_{\boldsymbol{u}\in\mathcal{U}}\big[L_{\boldsymbol{f}} V(\boldsymbol{x})+L_{\boldsymbol{g}} V(\boldsymbol{x})\boldsymbol{u}\big] \le -\alpha_V\!\left(V(\boldsymbol{x})\right), \qquad \forall \boldsymbol{x}\in\mathcal{X}.
\end{equation}
In the work presented here, $\alpha_V$ is restricted to the monotone class-$\mathcal{K}$ family, such that 
$\alpha_V(s) = c_V s$
with a scalar gain $c_V> 0$.

\subsection{Input-Constrained Control Barrier Functions}\label{ICCBFdetail}

 In RPO, thrust saturation must be respected explicitly, and many safety conditions are position-defined, thus often yielding a higher relative degree under second-order translational dynamics (i.e.,  $ L _ {\boldsymbol{g}} h = 0$ for a position-only barrier). ICCBFs address both these issues.

Under bounded actuation, the nominal safe set $\mathcal{S}=\{h\ge 0\}$ may not be forward invariant. ICCBFs construct an {input-admissible} inner safe set $\mathcal{C}^\star\subseteq\mathcal{S}$ by recursively composing $h$ with Lie derivatives and class-$\mathcal{K}$ functions \cite{agrawal2021safe}.
The ICCBF formulation is as follows. Define:
\begin{align}
b_0&\triangleq h(\boldsymbol{x}), \label{eq:iccbf-b0}\\
b_i&\triangleq \inf_{\boldsymbol{u}\in\mathcal{U}}\big[L_{\boldsymbol{f}} b_{i-1}(\boldsymbol{x})+L_{\boldsymbol{g}} b_{i-1}(\boldsymbol{x})\boldsymbol{u}+\alpha_{i-1}\!\left(b_{i-1}(\boldsymbol{x})\right)\big], \label{eq:iccbf-bi}
\end{align}
for $i=1,\dots,N$
with class-$\mathcal{K}$ functions $\{\alpha_i\}_{i=0}^{N-1}$. Let $\mathcal{C}_i\triangleq\{\boldsymbol{x}:b_i(\boldsymbol{x})\ge 0\}$ and
\begin{equation}\label{cstar}
\mathcal{C}^\star\triangleq \bigcap_{i=0}^{N}\mathcal{C}_i.
\end{equation}
If there exists a class-$\mathcal{K}$ function $\alpha_N$ such that
\begin{equation}
\sup_{\boldsymbol{u}\in\mathcal{U}}\big[L_{\boldsymbol{f}} b_N(\boldsymbol{x})+L_{\boldsymbol{g}} b_N(\boldsymbol{x})\boldsymbol{u}+\alpha_N\!\left(b_N(\boldsymbol{x})\right)\big]\ge 0,\qquad \forall \boldsymbol{x}\in\mathcal{C}^\star,
\end{equation}
then any locally Lipschitz feedback satisfying
\begin{equation}\label{eq:iccbf-constraint}
L_{\boldsymbol{f}} b_N(\boldsymbol{x})+L_{\boldsymbol{g}} b_N(\boldsymbol{x})\boldsymbol{u}+\alpha_N\!\left(b_N(\boldsymbol{x})\right)\ge 0
\end{equation}
renders $\mathcal{C}^\star$ forward invariant \cite{agrawal2021safe}.

\subsection{Meta-Reinforcement Learning}
Reinforcement learning (RL) addresses sequential decision-making problems in which an agent interacts with an environment and improves its behavior through trial and error \cite{Sutton1998}. Meta-RL extends RL from learning a single policy for a single task to learning an {adaptation mechanism} across a distribution of related tasks $\mathcal{M}\sim p(\mathcal{M})$ \cite{beck2025tutorial}. In this setting, individual tasks differ due to variations in parameters such as mass, thrust limits, or disturbance levels; thus, optimal behavior depends on properties that are not fully determined by instantaneous observations.

A policy $\pi_\theta$ may be represented either as a feed-forward mapping or as a recurrent dynamical system. In the feed-forward case, a multilayer perceptron (MLP) policy selects actions as $\boldsymbol{a}_k=\pi_\theta(\boldsymbol{o}_k)$ from the current observation $\boldsymbol{o}_k$, yielding a memoryless decision rule. In the recurrent case, an LSTM can be used to augment the mapping with an internal hidden state, i.e., $\boldsymbol{a}_k=\pi_\theta(\boldsymbol{o}_k,\boldsymbol{s}_k)$ with $\boldsymbol{s}_{k+1}=\phi_\theta(\boldsymbol{s}_k,\boldsymbol{o}_k,\boldsymbol{a}_k)$,  where $\boldsymbol{s}_k$ is the current hidden state and $\phi_\theta$ is the state update function of the hidden network, allowing the policy to integrate information over time. This internal memory enables within-episode inference of hidden task parameters and corresponding adaptation, making it well suited to meta-RL problems \cite{GAUDET2020180}.

\section{METHODOLOGY}\label{sec:method}
Given the control-affine system in Eq.~\eqref{xdoteq} with bounded input set $\mathcal{U}$ and a safety specification encoded by $h(\boldsymbol{x})\ge 0$, the goal of this work is to execute safe control under thrust saturation, time-sampled execution under ZOH, and significant parametric uncertainty and noise, while reducing fuel consumption and other myopic effects of CBF-based filtering.

The approach in this paper combines three ingredients: (1) an ICCBF-based inner safe set $\mathcal{C}^\star$ enforceable under bounded actuation, (2) a time-sampled margin designed to promote forward-invariance behavior under ZOH, and (3) a learned parameterization of the full class-$\mathcal{K}$ hierarchy within the ICCBF recursion, trained via meta-RL to improve feasibility and fuel usage across hidden parameters and uncertainties. At runtime, the control input is computed using a single convex QP per time step.

\subsection{ICCBF Recursion and Class-$\mathcal{K}$ Parameterization}
The ICCBF recursion in Section~\ref{ICCBFdetail} defines the sequence $\{b_i(x)\}_{i=0}^N$ and the inner safe set $\mathcal{C}^\star$ in \eqref{eq:iccbf-bi}--\eqref{cstar}. The set $\mathcal{C}^\star$ is strongly shaped by the choice of class-$\mathcal{K}$ functions $\{\alpha_i\}_{i=0}^{N}$. In this work, each $\alpha_i$ is parameterized as a monotone class-$\mathcal{K}$ function with positive gain,
\begin{equation}
\alpha_i(s_k) = \theta_{i,k}\, s_k,\qquad \theta_{i,k}>0,\qquad i=0,\dots,N,
\label{eq:alpha_linear}
\end{equation}
where the gains are {state-adaptive} and updated at each sampling instant $k$. Collecting the gains gives
\begin{equation}
\boldsymbol{\theta}_k \triangleq [\theta_{0,k},\theta_{1,k},\dots,\theta_{N,k}]^\top \in \mathbb{R}^{N+1}_{>0}.
\label{eq:theta}
\end{equation}
A policy $\pi_\psi$, trained via RL, maps the current state $x_k$ to the ICCBF parameters, i.e., $\boldsymbol{\theta}_k = \pi_\psi(x_k)$, thereby shaping the conservatism and feasibility of the inner safe set online. When a CLF is present, the policy additionally outputs its class-$\mathcal{K}$ parameter, i.e.,
$[\boldsymbol{\theta}_k, c_{V,k}] = \pi_\psi(x_k).$

\subsection{Control Margin Computation}\label{cmargin}
A time-sampled CBF must guarantee $\boldsymbol{x}(t)\in\mathcal{C}^\star$ for all $t\ge 0$, rather than only at the sampling instants.
To achieve this, a ZOH-ICCBF margin is added to the invariance condition at the sampled states, 
such that Eq.~\eqref{eq:iccbf-constraint} becomes 
\begin{equation}\label{eq8}
L_f b_N(\boldsymbol{x}) + L_g b_N(\boldsymbol{x}) u  + \alpha_N (b_N(\boldsymbol{x}))\geq   \nu(T,\boldsymbol{x})
\end{equation}
where $\nu(T,\boldsymbol{x})$ is the margin added. In this work, $\nu(T,\boldsymbol{x})$ is calculated using Theorem 1 in \cite{9417092}, where 
\begin{align}
 \nu(T,\boldsymbol{x}) &= l_1(\boldsymbol{x})T\Delta(\boldsymbol{x}) \\ 
    \Delta(\boldsymbol{x}) &=\sup_{\boldsymbol{z}\in\mathcal{X}(\boldsymbol{x},T),u\in\mathcal{U}}\|f(\boldsymbol{z})+g(\boldsymbol{z})u\| \label{nuvals} \\ 
    l_1(\boldsymbol{x}) &= l_{L_f b_N}(\boldsymbol{x}) + l_{L_g b_N}(\boldsymbol{x})u_{\max} + l_{\alpha(b_N)}(\boldsymbol{x}) \label{nuvals2}
\end{align}
where $l_{L_f b_N}$, $l_{L_g b_N}$, and $l_{{\alpha}(b_N)}$ are local Lipschitz constants for the set $\mathcal{R}(\boldsymbol{x},T)$. The proof of forward invariance under this margin can be found in \cite{9417092}. In \cite{9417092}, Eq.~\eqref{nuvals} and Eq.~\eqref{nuvals2} are computed via numerical maximization approximated by grid sampling. 
In contrast, this work computes these quantities using DA, yielding upper bounds with reduced computational burden.

At each sample time, a conservative local superset is defined as
$\mathcal{R}(\mathbf{x}_k,\zeta)\supseteq \mathcal{X}(\mathbf{x}_k,T)$, chosen as a hyper-rectangle centered at $\mathbf{x}_k$
with half-widths $\zeta\in\mathbb{R}^n_{>0}$:
\begin{equation}
\mathcal{R}(\mathbf{x}_k,\zeta)= \left\{\boldsymbol{x}\in\mathbb{R}^n : |x_i-(\mathbf{x}_k)_i|\le \zeta_i,\ i=1,\dots,n\right\}.
\label{eq:local_box}
\end{equation}
The Taylor approximations for all relevant quantities for the margin calculation are then constructed for the domain $\boldsymbol{x}=\mathbf{x}_k+\delta \mathbf{x}$ with $\delta x_i\in[-\zeta_i,\zeta_i]$.

For a continuously differentiable scalar function $p(x)$, a local Lipschitz constant on the domain
$\mathcal{R}(\mathbf{x}_k,\zeta)$ is defined as $l_{p}(\mathbf{x}_k)
\;\coloneqq\;
\sup_{x\in\mathcal{R}(\mathbf{x}_k,\zeta)}\|\nabla p(x)\|_2$
and satisfies
\begin{equation}
l_{p}(\mathbf{x}_k)
\leq 
\left(\sum_{i=1}^n \left(\sup_{x\in\mathcal{R}(\mathbf{x}_k,\zeta)}
\left|\frac{\partial p}{\partial x_i}(x)\right|\right)^2\right)^{1/2}.
\label{eq:lp_grad_bound}
\end{equation}

Using DA, $\partial p/\partial x_i$ can be computed by differentiation of the DA polynomial, and an interval enclosure for the Taylor approximation of each partial derivative can be obtained over $\mathcal{R}(\mathbf{x}_k,\zeta)$ using DA bounding. 

Let $\overline{d}_i$ denote the resulting upper bound
$\overline{d}_i \ge \sup_{x\in\mathcal{R}(\mathbf{x}_k,\zeta)}\big|\partial p/\partial x_i(x)\big|$.
Then a conservative bound for $l_{p}(\mathbf{x}_k)$ can be defined as 
\begin{equation}
\widehat{l}_{p}(\mathbf{x}_k)\;=\;
\left(\sum_{i=1}^n \overline{d}_i^2\right)^{1/2},
\label{eq:lp_hat}
\end{equation}
such that $\widehat{l}_{p}(\mathbf{x}_k)\ge {l}_{p}(\mathbf{x}_k)$. This can be utilized to calculate  conservative substitutes for $l_{L_{f}b_N}$, $l_{L_g b_N}$ and $l_{L_{\alpha(b_N)}}$.

Next, to bound $\Delta(\mathbf{x}_k)$, DA enclosures of the system dynamics are computed over $\mathcal{R}(\mathbf{x}_k,\zeta)$.
For control-affine dynamics in Eq.~\eqref{xdoteq} with admissible inputs
$\mathcal{U}=\{u:\|u\|_2\le u_{\max}\}$, the triangle inequality yields
\begin{align}
\sup_{\substack{x\in\mathcal{R}(\mathbf{x}_k,\zeta)\\ u\in\mathcal{U}}}\|f(x)+g(x)u\|_2
\;&\le\;
\sup_{x\in\mathcal{R}(\mathbf{x}_k,\zeta)}\|f(x)\|_2
\label{eq:delta_triangle_generic}\\
&\quad+\;
u_{\max} \sup_{x\in\mathcal{R}(\mathbf{x}_k,\zeta)}\|g(x)\|_2. \nonumber
\end{align}

where $\|\cdot\|_2$ denotes the induced matrix $2$-norm for $g(x)$.
Conservative values of the suprema in \eqref{eq:delta_triangle_generic} are obtained via DA interval enclosures
(component-wise bounds for $f$ and norm bounds for $g$), yielding a conservative value of $\Delta(\mathbf{x}_k)$.

Finally the resultant margin can be assembled as
\begin{align}\label{eq:nu_hat}
\widehat{l}_1(\mathbf{x}_k) &= \widehat{l}_{L_f b_N}(\mathbf{x}_k)
+\widehat{l}_{L_g b_N}(\mathbf{x}_k)\,u_{\max}
+\widehat{l}_{b_N}(\mathbf{x}_k),\\
\widehat{\nu}(T,\mathbf{x}_k) &= \widehat{l}_1(\mathbf{x}_k)\,T\,\widehat{\Delta}(\mathbf{x}_k).\label{eq:nu_hat2}
\end{align}
Since $\mathcal{X}(\mathbf{x}_k,T)\subseteq \mathcal{R}(\mathbf{x}_k,\zeta)$ and each DA bound is conservative over $\mathcal{R}(\mathbf{x}_k,\zeta)$ for the Taylor approximations, $\widehat{\nu}_1^\ell$ provides an adequate estimate for the margin required without needing any grid search methods to compute the Lipschitz constants.

However, note that the DA enclosures are computed for the truncated Taylor approximations rather than for the exact functions. For non-polynomial expressions, truncation introduces a modeling error that can, in principle, compromise conservatism. In this work, truncation effects are mitigated by selecting a sufficiently high DA expansion order and restricting the local domain so that the true function values lie within the obtained bounds. The obtained bounds are validated via extensive MC testing, as shown in the accompanying Jupyter notebooks. No bound violations were observed for any of the test cases. While Taylor models \cite{chachuat2012bounding} can be used to obtain bounds with formal guarantees for non-polynomial expressions, this would substantially increase computational complexity and is therefore beyond the scope of the present work due to the strong empirical validity of the DA-bounds observed.

\subsection{QP Formulations}\label{QPsection}
At each time step $k$, a QP is solved to compute a control input while satisfying the time-sampled ICCBF safety condition. The baseline per-step QP formulation is
\begin{equation}\label{eq:sf_iccbf}
\begin{aligned}
\mathbf{u}_k^*
&= \argmin_{\mathbf{u}_k \in \mathbb{R}^m}\|\mathbf{u}_k\|^2 \\
\text{s.t.}\;
& L_f b_N(\boldsymbol{x}_k)
  + L_g b_N(\boldsymbol{x}_k)\,\mathbf{u}_k \\
&\quad + \theta_{N,k} b_N(\boldsymbol{x}_k)
  \ge \widehat{\nu}_k\big(T,\boldsymbol{x}_k\big),
  \quad \mathbf{u}_k \in \mathcal{U}.
\end{aligned}
\end{equation}

where $b_N$ is the terminal ICCBF function, $\mathcal{K}_N>0$ is the terminal class-$\mathcal{K}$ gain, and $\widehat{\nu}_k$ is the time-sampled margin that accounts for inter-sample evolution under ZOH. 

For scenarios that also impose a convergence objective, a CLF constraint is added with a relaxation variable to preserve feasibility without compromising safety. The resulting CLF--ICCBF--QP is

\begin{equation}\label{eq:sf_clfcbf_qp}
\begin{aligned}
\mathbf{u}_k^*
&= \argmin_{\mathbf{u}_k \in \mathbb{R}^m,\;\varepsilon_k \ge 0}
   \tfrac{1}{2}\|\mathbf{u}_k\|^2 + p\,\varepsilon_k^2 \\
\text{s.t.}\;
& L_f b_N(\boldsymbol{x}_k)
  + L_g b_N(\boldsymbol{x}_k)\,\mathbf{u}_k \\
&\quad + \theta_{N,k}\, b_N(\boldsymbol{x}_k)
  \ge \widehat{\nu}_k\big(T,\boldsymbol{x}_k\big), \\
& L_f V(\boldsymbol{x}_k)
  + L_g V(\boldsymbol{x}_k)\,\mathbf{u}_k
  \le -c_{V,k}\,V(\boldsymbol{x}_k) + \varepsilon_k, \\
& \mathbf{u}_k \in \mathcal{U}.
\end{aligned}
\end{equation}

where $V$ is a CLF, $\alpha_V(\cdot)$ is a class-$\mathcal{K}$ function, and $p>0$ weights the CLF relaxation. The slack $\varepsilon$ relaxes the CLF constraint such that the CLF satisfaction does not come at the cost of the CBF, i.e at the expense of safety.  These optimizations are convex QPs as the ICCBF terminal condition is affine in $\mathbf{u}_k$. In all experiments conducted in this work, these QPs are solved using MOSEK \cite{mosek_apiremote_manual}.

\section{META-RL FORMULATION}\label{sec:learning}
The learning problem is posed as an episodic, discrete-time control task under time-sampled dynamics. At each time step $t_k$, the agent selects {ICCBF shaping parameters} $\boldsymbol{\theta}_k$ that are used by the QP formulation in Section \ref{QPsection} to compute the executed control input. The network architecture used, hidden parameter distributions,  and the RL environment are detailed below.

\subsection{Network Architecture}
\label{sec:net_arch}

\begin{figure}[tbp]
    \centering
    \includegraphics[width=\linewidth]{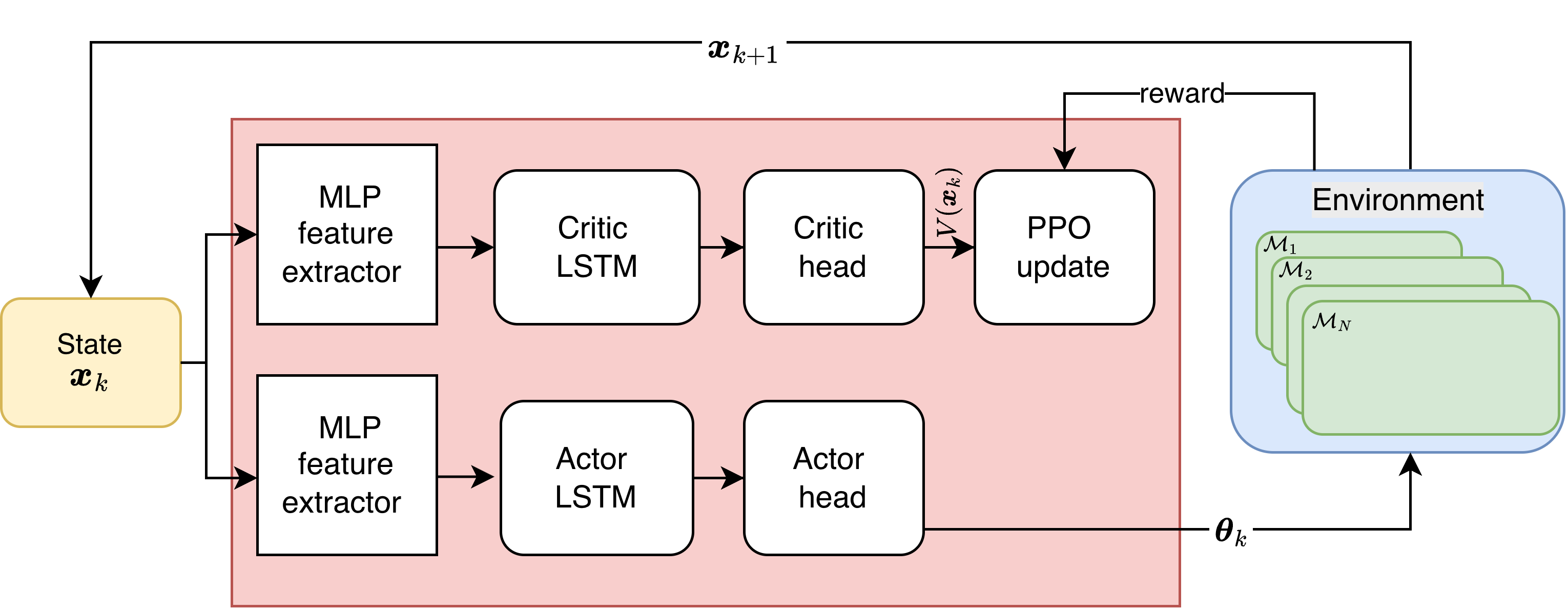}
    \caption{Meta-RL network architecture. State $x_k$ is mapped through separate actor/critic MLP feature extractors and separate LSTMs, producing ICCBF parameter outputs $\boldsymbol{\theta}_k$ from the actor head and value estimates $V(x_k)$ from the critic head. PPO updates are performed using rollouts across tasks $\mathcal{M}_i \sim p(\mathcal{M})$.}
    \label{netarc}
\end{figure}

Training uses a recurrent actor--critic policy implemented in Stable-Baselines3 \cite{raffin2021stablebaselines3}. The network architecture is shown in Fig.~\ref{netarc}. The observation is the full system state $x_k$. Separate feature extractors are used for the actor and critic along with independent recurrent modules  Thus, the actor processes $x_k$ through an actor-specific MLP feature extractor and produces the ICCBF shaping parameters $\boldsymbol{\theta}_k$, which are scaled to satisfy $\boldsymbol{\theta}_k \in [\boldsymbol{\theta}_{\min},\,\boldsymbol{\theta}_{\max}]$. The critic employs a separate MLP feature extractor followed by an LSTM to integrate temporal information and infer knowledge of the hidden parameters relevant for value estimation under task variation, and outputs a value function $V(x_k)$ for advantage computation, which is utilized in the policy update. Policy parameters are optimized using Proximal Policy Optimization (PPO) \cite{schulman2017proximal,WIJAYATUNGA2025109996}.

\subsection{Initial State and Hidden Parameter Distributions}

Each episode is initialized by sampling an initial condition and several variable environment parameters, such as mass, maximum thrust $u_{\max}$, size of KIZ/KOZ, etc.  
This episode-level randomization exposes the meta-RL agent to a distribution of task instances, improving robustness and reducing overfitting to a single nominal scenario. 

At the start of an episode, the environment draws an initial state $x_0$ from a bounded admissible set $\mathcal{X}_0 \in \mathcal{S}$.
Each episode also samples a vector of hidden parameters
$\mathbf{p} \in \mathbb{R}^{n_p}$ from a bounded hyper-rectangle centred about nominal values
$ \mathbf{\bar{p}}$. Each parameter $p_i$ is drawn independently as
\begin{equation}\label{eq:drawparam}
p_i \sim \mathrm{Uniform}\!\left(\bigl[(1-\delta_i)\,\bar p_i,\ (1+\delta_i)\,\bar p_i\bigr]\right),
\end{equation}
for $i = 1,\dots,n_p$, where $\bar p_i$ denotes the nominal value of the $i$th parameter and $\delta_i > 0$
specifies its relative variation level. The percentage variations $\delta_i$ used for each test case presented in the results are summarized in Table~\ref{tab:uncertainty_summary}.
In the MC results provided later, the parameter samples are held fixed across controllers within each dataset to ensure that performance differences arise solely from the proposed ICCBF tuning mechanism.

\subsection{RL environment}
As mentioned, PPO is used in this work to train a neural network that outputs the ICCBF class-$\mathcal{K}$ gains $\boldsymbol{\theta}_k$ and also optionally the CLF gain $c_{V,k}$ in Eq.~\eqref{eq:sf_clfcbf_qp} at timestep $k$. Once the actor provides these the ICCBF-QP formulation in Eq.~\eqref{eq:sf_iccbf} or \eqref{eq:sf_clfcbf_qp} can be used to calculate the control $\mathbf{u}_k$ needed to retain safety till the next time step while also (optionally) satisfying the CLF condition.  The state is then propagated
over one sampling period $T$ under zero-order hold using \eqref{xdoteq} to obtain $\mathbf{x}_{k+1}$.
 Algorithm~\ref{alg:env_step} summarizes this process.

\begin{algorithm}[tbp]
\caption{State Transition \& Reward Computation}
\label{alg:env_step}
\begin{algorithmic}[1]
\Require $T$, $t$, $t_f$, state $\mathbf{x}_k$, $\sigma_{\parallel \mathbf{u}\parallel}$, $\sigma_{\mathbf{u}_\theta}$,
$\sigma_\mathbf{x}$, $\theta_k$, (optional) $\alpha'$
\State Update time $t \gets t + T$.
\State Set $\texttt{done} = \textbf{false}$ and $\texttt{failure} = \textbf{false}$.
\State Calculate $\widehat{\nu}
(T,\mathbf{x}_k)$ using \eqref{eq:nu_hat} and \eqref{eq:nu_hat2}.
\State Construct the ICCBF--QP as shown in \eqref{eq:sf_clfcbf_qp}.
\State Solve \eqref{eq:sf_clfcbf_qp} to obtain nominal control $\mathbf{u}_k^\star$.
\If{QP infeasible or solver failure}
    \State Set \texttt{done} $\gets$ \textbf{true}; \texttt{failure} $\gets$ \textbf{true}; $\mathbf{u}_k^\star \gets \mathbf{0}$.
\EndIf
\State {Apply control noise as shown in Eq.~\eqref{eq:execution_noise_model} to obtain $\mathbf{u}_k$}.
\State Propagate the state with $\mathbf{u}_k$ over $T$ using Eq.~\eqref{xdoteq} to obtain $x_{k+1}$.
\State Compute the original safety function value $h_{k+1}$ and (if available) CLF value $V_{k+1}$.
\State Add noise to form the observation $\pmb{S}_k$ using Eq.~\eqref{eq:statenoise} and scale using Eq.~\eqref{scaleobs} to obtain $\pmb{S}_{k+1}^*$.

\If{$t \geq t_f$}
    \State $\texttt{done}\gets \textbf{true}$.
\EndIf
\If{$h_{k+1} < 0$}
    \State $\texttt{done}\gets \textbf{true}$, $\texttt{failure}\gets \textbf{true}$.
\EndIf
\State Compute reward $R_k$ using Eq.~\eqref{eq:Rk_def}.
\State \Return $(\pmb{S}_{k+1}, R_k, \texttt{done})$.
\end{algorithmic}
\end{algorithm}

\paragraph{State Errors}

State errors are considered in this work, and are modeled as additive zero-mean Gaussian noise on the true state. At each time step $k$, the state available to the learning agent is calculated as
\begin{equation}\label{eq:statenoise}
    \boldsymbol{x}_k^E = \boldsymbol{x}_k + \boldsymbol{\epsilon}_{\boldsymbol{x}_k},
    \qquad
    \boldsymbol{\epsilon}_{\boldsymbol{x}_k} \sim \mathcal{N}\!\left(\boldsymbol{0},\,
    \boldsymbol{\sigma}_x^2 \mathbf{I}\right),
\end{equation}
where $\boldsymbol{\sigma}_x = [\boldsymbol{\sigma}_r, \boldsymbol{\sigma}_v]^T$ denotes the standard deviation of the position and velocity errors.

\paragraph{Thrust Errors}
To model thrust execution errors, the applied control $\mathbf{u}_k$ is obtained by perturbing the magnitude
and direction of $\mathbf{u}_k^\star$ prior to propagation. Let $\mathbf{u}_k^\star=[u_x,u_y,u_z]^\top$ be
expressed in the same control frame used by the dynamics. Define the magnitude
$u_k=\|\mathbf{u}_k^\star\|_2$ and the out-of-plane and in-plane angles $(\beta,\gamma)$, then
apply independent magnitude and angular perturbations and reconstruct the executed command as
\begin{align}\label{eq:execution_noise_model}
u_k &= \|\mathbf{u}_k^\star\|_2,\qquad
\beta = \sin^{-1}\!\left(\frac{u_z}{u_k}\right),\qquad
\gamma = \tan^{-1}\!\left(\frac{u_x}{u_y}\right), \\
u_{k}^{E} &= u_{k}\left(1+\delta_u\right),\qquad
\beta^{E} = \beta + \delta_{\beta},\qquad
\gamma^{E} = \gamma + \delta_{\gamma}, \\
\mathbf{u}_k &= u_{k}^{E}
\begin{bmatrix}
\cos \beta^{E}\,\sin \gamma^{E}\\
\cos \beta^{E}\,\cos \gamma^{E}\\
\sin \beta^{E}
\end{bmatrix},
\qquad
\mathbf{u}_k \leftarrow \frac{u_{\max}}{\max(u_{\max},\|\mathbf{u}_k\|_2)}\,\mathbf{u}_k.
\end{align}

where $\delta_u \sim \mathcal{N}(0,\sigma_u^2)$, $\delta_{\beta} \sim \mathcal{N}(0,\sigma_\beta^2)$, and
$\delta_{\gamma} \sim \mathcal{N}(0,\sigma_\gamma^2)$.

\paragraph{Observation}
At each time step, the RL environment must also provide an observation $\pmb{S}_k$ that contains sufficient information to estimate the potential reward of an action at each step $k$.
To normalize inputs for stability and consistent feature scaling, a simple min-max normalization is used such that 
\begin{equation}\label{scaleobs}
\pmb{S}_k^*=2\left(\frac{\pmb{S}_k-\pmb{S}_{\min}}{\pmb{S}_{m a x}-\pmb{S}_{\min}}\right)-1.
\end{equation}
{where $\pmb{S}_{\min}$ and $\pmb{S}_{\max}$  are the vectors of minimum and maximum values of the components in $\pmb{S}$, respectively. 

For the cruise control and docking cases considered, $\pmb{S}_k = \boldsymbol{x}_k^E$. For the inspection case, $\pmb{S}_k = [\boldsymbol{x}_k^E, \theta_S,  n_{insp}, \hat{\mathbf{d}}]^T$ where 
$\theta_S$ is the sun-angle, $n_{insp}$ is the number of points inspected, and $\hat{d}$ is the direction vector that points to the largest cluster of remaining uninspected points, computed using K-means as described in \cite{doi:10.2514/1.I011391}.
\paragraph{Reward}
The per-step reward penalises control effort and failure events, with an additional terminal CLF penalty at
the episode horizon. Specifically,
\begin{equation}
R_k=
\begin{cases}
r_k, & t<t_f,\\
r_k-w_V P_V, & t\ge t_f,
\end{cases}
\label{eq:Rk_def}
\end{equation}
where
\begin{equation}\label{CBFreward}
r_k
=
- w_u \lVert \mathbf{u}_k^\star \rVert_2
- w_{\mathrm{fail}} P_{\mathrm{fail}}
- w_h \, \max\!\bigl(0,\,-h_{k+1}\bigr).
\end{equation}
Here, $w_u$, $w_{\mathrm{fail}}$, $w_h$, and $w_V$ are user-selected weights, and $P_{\mathrm{fail}}$ flags
QP failures:
\begin{equation}
P_{\mathrm{fail}}=
\begin{cases}
1, & \text{QP infeasible or solver failure},\\
0, & \text{otherwise}.
\end{cases}
\label{eq:Pfail_def}
\end{equation}
When a CLF is used, a terminal penalty is applied if the CLF fails to decrease below a prescribed threshold
$\rho_V$ over the episode. Let $N_f=\lceil t_f/T\rceil$ and $\mathbf{V}=\{V(\mathbf{x}_j)\}_{j=0}^{N_f}$. Then
\begin{equation}
P_V=
\begin{cases}
\min(\mathbf{V}), & \min(\mathbf{V})>\rho_V,\\
0, & \text{otherwise}.
\end{cases}
\label{eq:PV_def}
\end{equation}



\section{RESULTS}\label{sec:experiments}

In this section, three test cases of increasing dimension are considered: (i) 1D cruise control, (ii) 2D spacecraft docking with a rotating target, both adapted from \cite{agrawal2021safe}, and (iii) a 3D inspection scenario adapted from \cite{doi:10.2514/1.I011391}. An episode is marked {safe} if all safety constraints remain satisfied for the full horizon or till the mission goal is achieved.  Fuel usage is reported using either the total fuel consumption over time $\sum_k \|\mathbf{u}_k\|T$ or total $\Delta v$ ($/m\sum_k \|\mathbf{u}_k\|T$). Fuel consumption results (and inspection metric for the inspection problem ) are reported in terms of mean $\mu$ $\pm$ standard deviation $\sigma$, together with quartiles [$Q_1, Q_2, Q_3, Q_4$] and the 99th percentile $P_{99}$.

Three controllers are compared for each case: (i) an \emph{untuned} ICCBF baseline using constant [$\boldsymbol{\theta}_k$,$c_{V,k}$] values, (ii) an \emph{MLP-tuned} ICCBF in which a feed-forward PPO policy outputs [$\boldsymbol{\theta}_k$,$c_{V,k}$], and (iii) the \emph{RNN-tuned} ICCBF in which a recurrent PPO policy using an LSTM actor-critic outputs [$\boldsymbol{\theta}_k$,$c_{V,k}$].   This section details the evaluation protocol, the hyperparameters used for training, and the details and results of the three test cases.

Training and network hyperparameters are summarized in Table~\ref{tab:hyperparams_main}. Hyperparameters were iteratively tuned to ensure stable training across tasks, and identical PPO settings were used across baselines in each test case.

\begin{table}[t]
\centering
\caption{Main hyperparameters for the three test cases.}
\label{tab:hyperparams_main}
\begin{tabular}{lrrr}
\hline
\textbf{Hyperparameter} & \textbf{Cruise Control} & \textbf{Docking} & \textbf{Inspection} \\ \hline
\multicolumn{4}{l}{\textit{Environment / rollout}}                                                                                 \\
Time step $T$           & 0.1                           & 0.5                    & 10                                              \\
Total timesteps         & $10^5$                        & $10^6$                 & $10^6$                                          \\
Batch size              & 64                            & 64                     & 256                                              \\
Epochs per update       & 10                            & 10                     & 10                                              \\ \hline
\multicolumn{4}{l}{\textit{Optimization / PPO settings}}                                                                           \\
Learning rate           & $10^{-4}$                     & $10^{-4}$              & $2\times 10^{-4}$                               \\
Discount factor         & 0.99                          & 0.995                  & 0.99                                            \\
GAE                     & 0.95                          & 0.95                   & 0.95                                            \\
Clip range              & 0.10                          & 0.10                   & 0.20                                            \\
Entropy coef.           & 0.01                          & 0.01                   & 0.01                                            \\ \hline
\multicolumn{4}{l}{\textit{Network architecture (shared across actor/critic unless noted)}}                                        \\
Activation              & \texttt{Tanh}                  & \texttt{Tanh}           & \texttt{Tanh}                                   \\
Hidden layers           & 3                             & 4                      & 4                                               \\
Nodes per layer         & 32                            & 64                     & 128                                             \\
std                     & 0.2                           & 0.2                    & 0.2                                             \\ \hline
\multicolumn{4}{l}{\textit{LSTM (RNN only)}}                                                                         \\
LSTM hidden size        & 64                            & 64                     & 128                                             \\
\# LSTM layers          & 1                             & 1                      & 1                                               \\ \hline
\end{tabular}
\end{table}

To ensure fair comparisons across learned policies, for each test case evaluation is performed on a fixed MC dataset of $N_{\mathrm{MC}}$ episodes,
$\mathcal{D}_{\mathrm{MC}} \triangleq \{(\mathbf{x}_0^{(i)},\mathbf{p}^{(i)})\}_{i=1}^{N_{\mathrm{MC}}}$
where initial conditions and episode parameters are pre-sampled and held fixed across all methods. 

In all test cases discussed, the ICCBF order $N$ is set to 2, which is the minimal ICCBF order required for position-defined constraints under the second-order translational dynamics considered. For the cruise-control task, where the safety constraint has relative degree one, $N=2$ is not strictly necessary but is retained for consistency and to allow additional shaping of the inner safe set.


\begin{table*}[t]
\centering
\caption{Hidden parameter variations per episode (nominal value with relative deviation $\delta_i$)}
\label{tab:uncertainty_summary}
\begin{tabular}{lrrr}
\hline
\textbf{Category} & \textbf{Cruise Control} & \textbf{Docking} & \textbf{Inspection} \\
\hline
Deputy mass $m$ 
& $1650~\mathrm{kg}$ ($\pm 20\%$) 
& $1000~\mathrm{kg}$ ($\pm 10\%$) 
& $12~\mathrm{kg}$ ($\pm 10\%$) \\

Max thrust $u_{\max}$ 
& $0.25~\mathrm{N}$ ($\pm 20\%$) 
& $0.25~\mathrm{kN}$ ($\pm 10\%$) 
& $1.0~\mathrm{N}$ ($\pm 10\%$) \\

Max velocity $v_{\max}$ 
& $24.0~\mathrm{m\,s^{-1}}$ ($\pm 20\%$) 
& -- 
& -- \\

Lead vehicle velocity $v_0$ 
& $13.89~\mathrm{m\,s^{-1}}$ ($\pm 10\%$) 
& -- 
& -- \\

Deputy radius $R_D$ 
& -- 
& -- 
& $5~\mathrm{m}$ ($\pm 10\%$) \\

Chief radius $R_C$ 
& -- 
& $2.4~\mathrm{m}$ ($\pm 10\%$) 
& $10~\mathrm{m}$ ($\pm 10\%$) \\

Chief spin rate $\omega$ 
& -- 
& $0.6^\circ\,\mathrm{s^{-1}}$ ($\pm 10\%$) 
& -- \\

Chief orbit radius $r$ 
& -- 
& $6771~\mathrm{km}$ ($\pm 10\%$) 
& $6771~\mathrm{km}$ ($\pm 10\%$) \\

Approach cone half-angle $\gamma$ 
& -- 
& $10^\circ$ ($\pm 10\%$) 
& -- \\

KIZ radius $R_{\max}$ 
& -- 
& -- 
& $800~\mathrm{m}$ ($\pm 10\%$) \\

\hline
\end{tabular}
\end{table*}

\begin{table}[t]
\centering
\caption{Standard deviations of thrust and state uncertainties applied}
\label{tab:noise_models}
\begin{tabular}{lrrr}
\hline
\textbf{Noise source} & \textbf{Cruise Control} & \textbf{Docking} & \textbf{Inspection} \\ \hline 
$\sigma_u$
& $0.1$ 
& $0.05$ 
& $0.05$ \\
$\sigma_\gamma$
& -- 
& $0.1^\circ$ 
& $ 0.1^\circ$ \\
$\sigma_\beta$
& --  & --  
& $0.1^\circ$ \\
${\sigma}_r$ (each axis)
& $2~\mathrm{m}$ 
& $0.1~\mathrm{m}$ 
& $0.1~\mathrm{m}$ \\
${\sigma}_v$ (each axis)
& $0.5~\mathrm{m\,s^{-1}}$ 
& $2~\mathrm{mm\,s^{-1}}$ 
& $2~\mathrm{mm\,s^{-1}}$ \\
\hline
\end{tabular}
\end{table}

\begin{table*}[t]
\centering
\footnotesize
\setlength{\tabcolsep}{5pt}
\renewcommand{\arraystretch}{1.05}
\caption{Performance summary across tasks}
\label{tab:performance_summary}
\begin{tabular}{@{}l l r r c@{}}
\hline
\multirow{2}{*}{\textbf{Task}} &
\multirow{2}{*}{\textbf{Case}} &
 $\boldsymbol{\mu \pm \sigma}$ & $\boldsymbol{[Q_1,\,Q_2,\,Q_3,\,P_{99}]}$ & 
\multirow{2}{*}{\textbf{\begin{tabular}[c]{@{}c@{}}Safe\\ $N_{\mathrm{safe}}/N_{\mathrm{MC}}$ (\%)\end{tabular}}} \\
\cline{3-4}
& \multicolumn{1}{l}{\textbf{}} & \multicolumn{2}{c}{\textbf\textbf{Fuel consumption [Ns for Cruise Control, m/s for others]}} & \\
\hline
\multirow{3}{*}{Cruise Control}
& Untuned   & $4.75 \pm 0.74$ & $[4.26,\,4.77,\,5.33,\,5.96]$ & 99.82\% \\
& MLP-Tuned & $3.86 \pm 0.51$ & $[3.53,\,3.86,\,4.21,\,4.91]$ & 99.82\% \\
& RNN-Tuned & $2.77 \pm 0.62$ & $[2.27,\,2.74,\,3.24,\,4.08]$ & 99.84\% \\
\hline
\multirow{3}{*}{Docking}
& Untuned   & $8.20 \pm 1.94$ & $[7.35,\,8.05,\,9.30,\,12.00]$ & 96.88\% \\
& MLP-Tuned & $5.70 \pm 1.79$ & $[4.45,\,5.77,\,6.79,\,11.09]$ & 97.06\% \\
& RNN-Tuned & $6.19 \pm 1.51$ & $[5.51,\,6.15,\,7.00,\,10.09]$ & 97.28\% \\
\hline
\multirow{6}{*}{Inspection}
& Untuned   & $289.32 \pm 194.89$ & $[109.54,\,297.13,\,429.18,\,866.62]$ & 100\% \\
& MLP-Tuned & $27.39 \pm 74.98$   & $[7.83,\,8.50,\,9.55,\,392.12]$       & 88.4\% \\
& RNN-Tuned & $3.87 \pm 5.53$     & $[0.84,\,1.10,\,2.00,\,15.27]$         & 100\% \\
\cline{3-4}
& \multicolumn{1}{l}{\textbf{}} & \multicolumn{2}{c}{{Inspection Score (\%)}} & \\
\cline{3-4}
& Untuned   & $99.99 \pm 0.14$  & $[100.00,\,100.00,\,100.00,\,100.00]$ & \\
& MLP-Tuned & $31.59 \pm 14.51$ & $[24.00,\,26.00,\,33.00,\,88.50]$     & \\
& RNN-Tuned & $85.41 \pm 27.25$ & $[100.00,\,100.00,\,100.00,\,100.00]$ & \\
\hline
\end{tabular}
\end{table*}

\subsection{Cruise Control Problem}\label{ccproblemintro}

Discussed in \cite{7040372} and \cite{agrawal2021safe}, this problem considers a point-mass model of a vehicle (chief) moving along a straight line. A following vehicle (deputy) trails a lead vehicle at a distance $d$, with the lead vehicle travelling at a known constant speed $v_0$. The objective is to design a controller that drives the deputy to the speed limit $v_{max}$ while ensuring collision avoidance. The collision avoidance safety constraint is specified as $d \geq 1.8v$, and the CLF constraint that drives the following vehicle to the speed limit
is defined as $V(x)  =(v - v_{\text{max}})^2$. Defining the state vector as $x = [d,\, v]^T$, the dynamical model is
\begin{equation}
\begin{bmatrix}
\dot{d} \\
\dot{v}
\end{bmatrix}
=
\begin{bmatrix}
v_0 - v \\
-\tfrac{F(v)}{m}
\end{bmatrix}
+
\begin{bmatrix}
0 \\
g_0
\end{bmatrix} u,
\qquad
\mathcal{U} = \{u : |u| \leq u_{\max}\},
\end{equation}
where $u$ is the control input. The resistive force $F(v)$ is modeled as $F(v) = 0.1 + 5v + 0.25 v^2$, $m$ is the deputy mass, and $g_0$ is the gravitational acceleration.  The safe set $\mathcal{S}$ is then defined as
\begin{equation}
\mathcal{S} = \{x \in \mathcal{X} \;|\; h_0(x) = x_1 - 1.8x_2 \geq 0\}.
\end{equation}
The initial states are drawn such that $\mathbf{x}_0 \in \mathcal{S}$. The hidden parameter vector
includes the vehicle mass $m$, the front-vehicle speed $v_0$, the target speed $v_{\max}$, and the maximum
available thrust $u_{\max}$. 
At the start of each episode, the hidden parameters are sampled according to Eq.~\eqref{eq:drawparam}, with the nominal values and deviations given in Table \ref{tab:uncertainty_summary}. For the 5000 MC samples analyzed, the drawn hidden parameters are shown in Figure \ref{fig:cc2}. The standard deviation of the position and velocity uncertainties added is given in Table \ref{tab:noise_models}.

\begin{figure}[tbp]
    \centering
    \includegraphics[width=\linewidth]{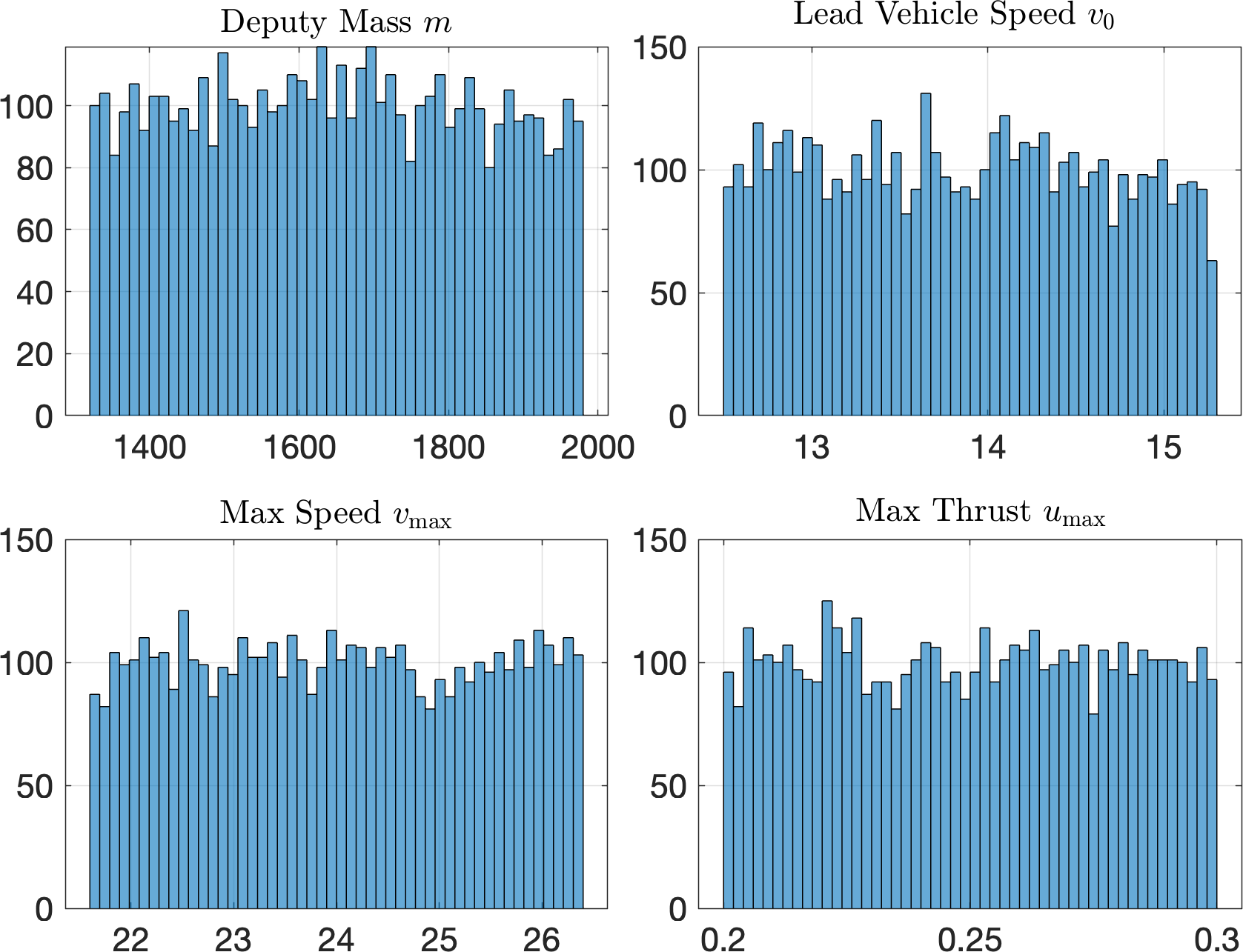}
        \caption{Cruise Control MC parameter variations in the 5000 samples}
    \label{fig:cc2}
\end{figure}

\paragraph*{Outcomes}
Fig.~\ref{fig:cc1} and Table~\ref{tab:performance_summary} summarize the thrust consumption statistics for the cruise-control task. The untuned ICCBF ($\boldsymbol{\theta} = [4,7, 2]$ and $\alpha_V = 10$) exhibits a comparatively high thrust demand, with a heavy upper tail. 
Tuning the ICCBF is seen to substantially reduce thrust consumption. The MLP-tuned ICCBF decreases the mean thrust by $ 18.7\%$ relative to the untuned ICCBF. While the RNN-tuned ICCBF achieves the largest fuel savings, providing a $41.7\%$ reduction relative to the untuned ICCBF and an $28.2\%$ reduction relative to the MLP-tuned ICCBF. The reduction is also reflected in the median and interquartile statistics, with the upper quartile $Q_3$ of RNN-tuned ICCBF showing a $39.2\%$. The upper tail is significantly compressed as well with the $P_{99}$ reduced by $31.5\%$.


\begin{figure}[tpb]
    \centering    \includegraphics[width=\linewidth]{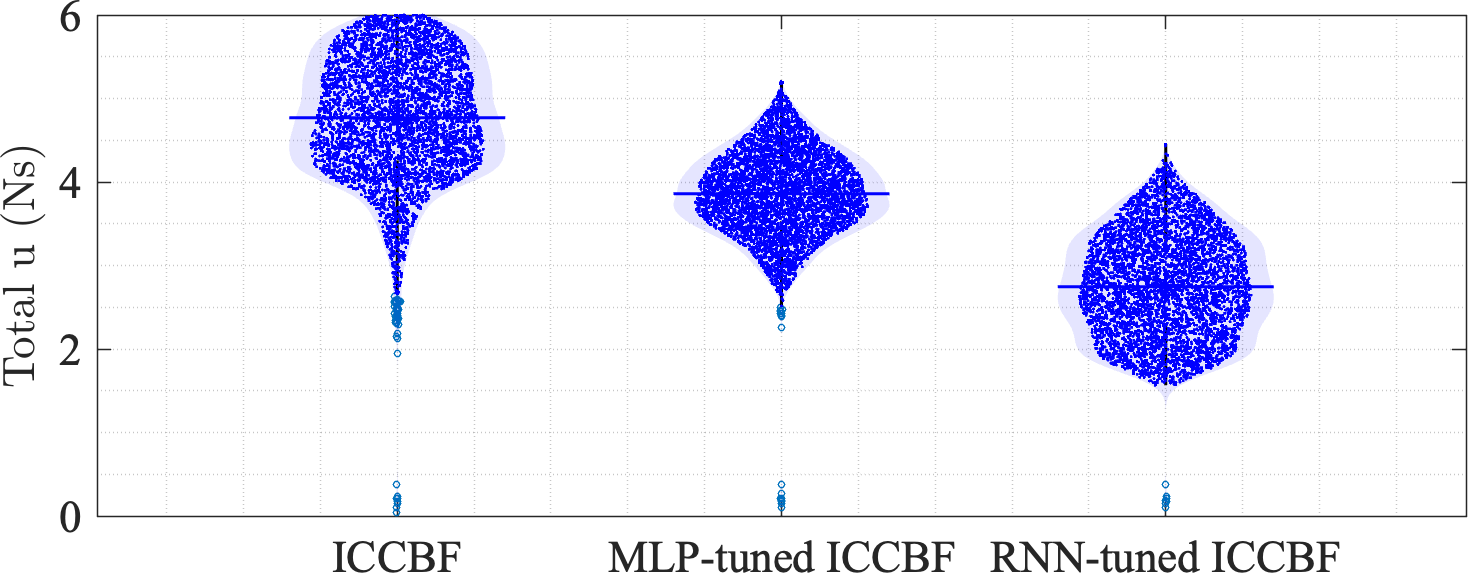}
        \caption{Cruise control total thrust distributions}
    \label{fig:cc1}
\end{figure}


Figure \ref{fig:ccres} illustrates a sample of the resultant MC trajectories, along with the variation of the CLF and CBF conditions and the number of failure cases observed. All three methods are seen to exhibit comparably low failure counts (on the order of $\sim 0.2\%$), with the RNN-tuned ICCBF slightly improving the empirical success rate. 

\begin{figure}[tbp]
    \centering
    \includegraphics[width=\linewidth]{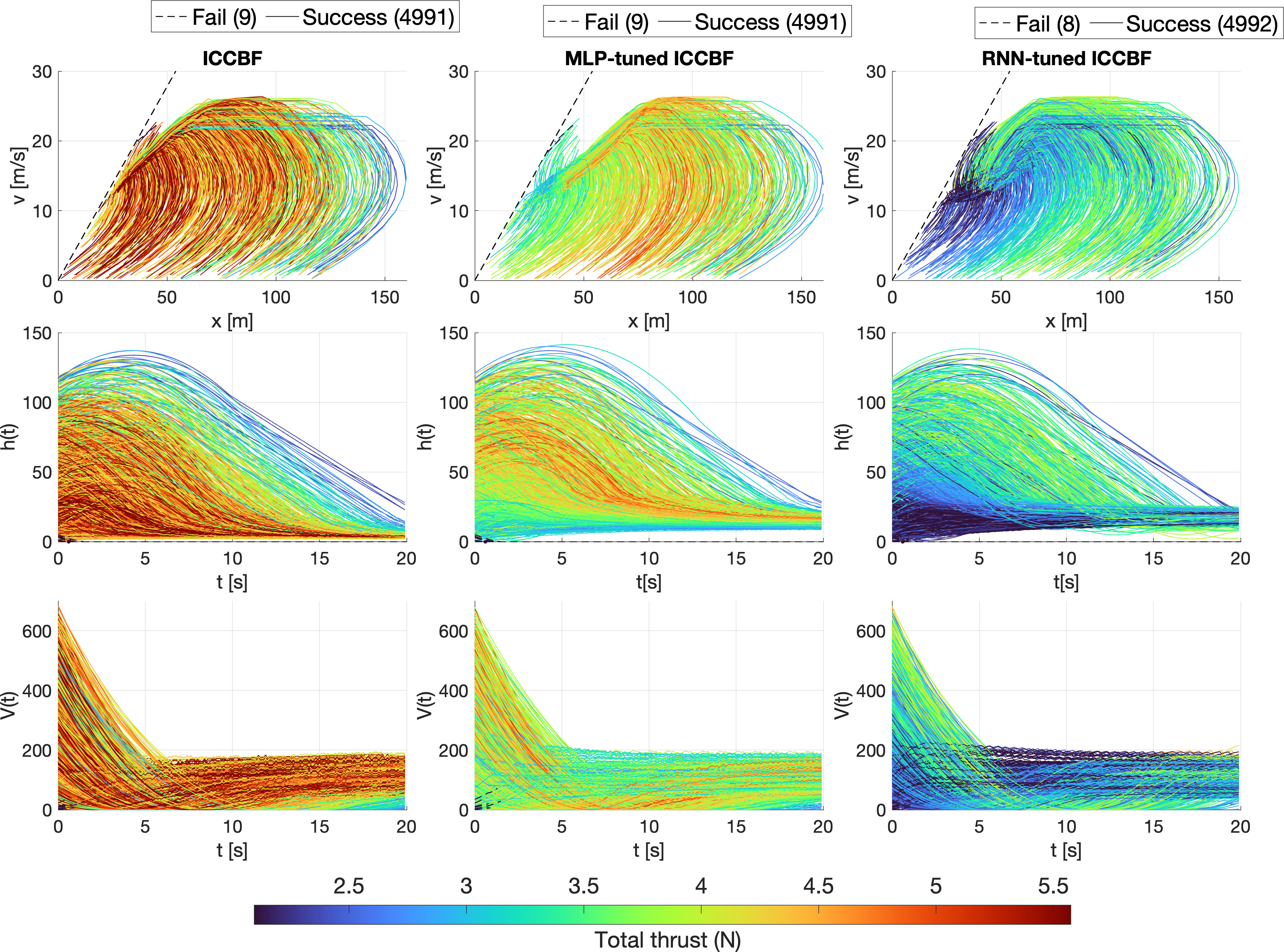}
    \caption{Cruise Control trajectories and CBF, CLF variation over time}
    \label{fig:ccres}
\end{figure}

\subsection{Spacecraft Rendezvous with Rotating Target}
\begin{figure}[tbp]
\centering\includegraphics[width=\linewidth]{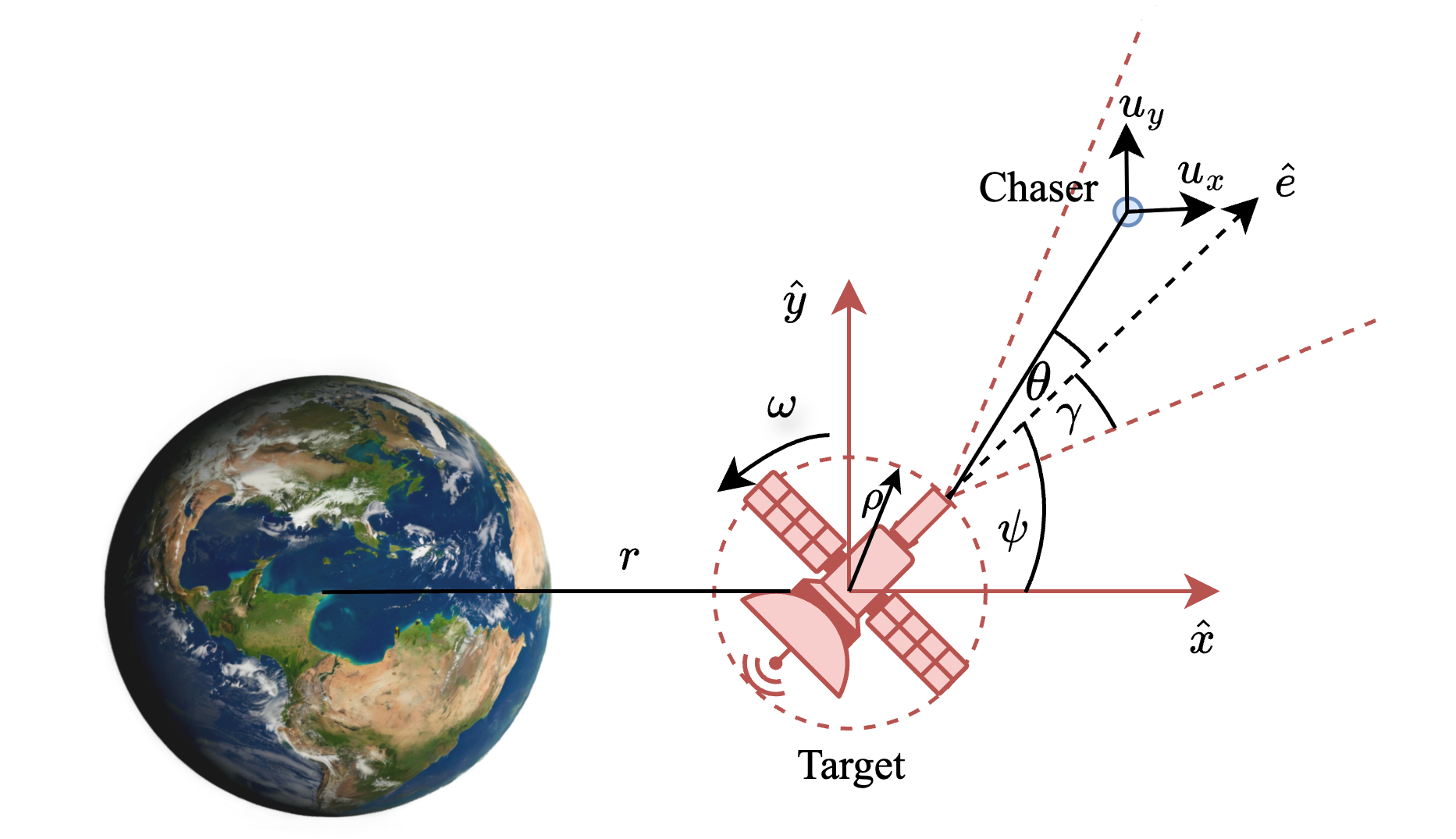}
    \caption{Spacecraft Docking Problem}
    \label{dockingproblemFig}
\end{figure}
Discussed in \cite{agrawal2021safe} and \cite{5991151}, this test case is an autonomous rendezvous scenario between a chaser spacecraft and a target body, both modeled as point masses. The target is represented as a point on a disk of radius $R_C$, rotating with a constant angular velocity $\omega$ relative to the Local-Vertical Local-Horizontal (LVLH) frame. The objective is to determine the thrust needed to bring the chaser spacecraft from an initial range of 100~m to within 3~m of the target. This test case is illustrated in Fig.~\ref{fig:dockingdiag}.

A LOS safety constraint is imposed in this test case, requiring that the chaser’s relative position remain inside a cone of half-angle $\gamma$ aligned with the docking axis at all times. The system state is defined as $x \in \mathbb{R}^5$, comprising the relative position $(p_x, p_y)$, relative velocity $(v_x, v_y)$, and the docking port angle $\psi$. The system dyanmics are nonlinear relative equations of motion, given in Eq. in \cite{agrawal2017discrete}.  The control inputs $\boldsymbol{u} = [u_x, u_y]^T$ represent the propulsive forces and are bounded such that $\parallel \boldsymbol{u}\parallel  \leq u_{\max}$.   The LOS constraint is expressed as $h_0(x) \geq 0$, where 
\begin{equation}
    h_0(x) = \cos\theta - \cos\gamma = \frac{\vec{r}_{c-p} \cdot \hat{e}}{\|\vec{r}_{c-p}\|} - \cos(\gamma)
\end{equation}
Here, $\vec{r}_{c-p} = \big(p_x - R_C \cos\psi, \; p_y - R_C \sin\psi\big)^T$ is the position vector of the chaser relative to the docking port, and $\hat{e} = (\cos\psi, \sin\psi)^T$ is the docking axis unit vector.  Note that this $h_0(x)$ has relative degree challenges as $L_g h_0(x) = 0$. To guide the chaser to the docking port the following CLF is used 
\begin{equation}
V(x) = 
\left( v_x + \frac{p_x - R_C \cos\psi}{10} \right)^2
+ \left( v_y + \frac{p_y - R_C \sin\psi}{10} \right)^2.
\end{equation}

In this case, the initial states are again drawn such that $\mathbf{x}_0 \in \mathcal{S}$. The episode parameter vector
includes the deputy mass $m$, chaser radius $R_C$, rotational rate of the target $\omega$, and the maximum
available thrust $u_{\max}$.
At the start of each episode, the parameters are sampled according to \eqref{eq:drawparam}, with nominal values and deviations given in Table~\ref{tab:uncertainty_summary}
The standard deviation of the state and thrust uncertainties added are given in Table~\ref{tab:noise_models}.

\begin{figure}[tbp]
    \centering
\includegraphics[width=\linewidth]{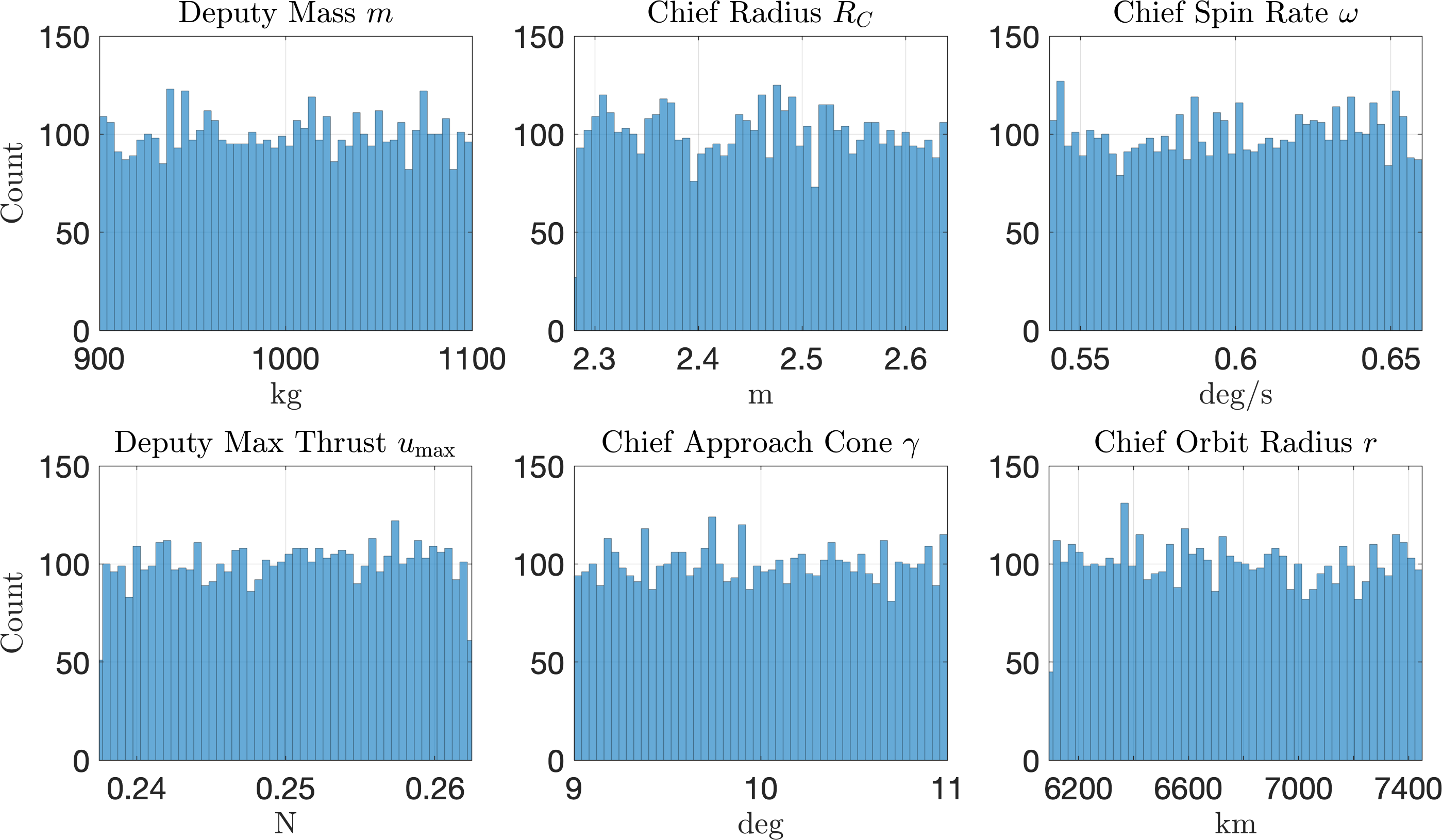}
    \caption{MC parameter variations for docking 5000 samples}
    \label{fig:dockingdiag}
\end{figure}




\subsubsection*{Outcomes}

Fig.~\ref{fig:dock}, Fig.~\ref{fig:dock2}, and Table~\ref{tab:performance_summary} summarize the $\Delta v$ consumption and safety performance for the docking task. Relative to the untuned ICCBF baseline ($\boldsymbol{\theta} = [0.25,0.85, 0.05]$ and $\alpha_V = 0.1$), both learned tunings reduce control efforts and minimize failure cases. The MLP-tuned ICCBF achieves the lowest mean total $\Delta v$, reducing the mean from $\SI{8.20}{\meter\per\second}$ to $\SI{5.70}{\meter\per\second}$, corresponding to a $30.3\%$ reduction from untuned ICCBF. The RNN-tuned ICCBF also yields a substantial reduction in thrust consumption, with a mean of $\SI{6.19}{\meter\per\second}$ ($24.5\%$
reduction). These improvements are consistently reflected in the median and interquartile statistics.
Although the RNN-tuned ICCBF doesn't have the minimum mean $\Delta v$, it exhibits a smaller upper tail with a lower $P_{99}$ value than the MLP-tuned ICCBF. This indicates improved robustness against high-thrust outliers.

All three controllers exhibit low failure fractions across the MC set. The untuned ICCBF yields a $1.86\%$ failure rate, while the MLP- and RNN-tuned ICCBFs yield $1.80\%$ and $1.74\%$, respectively. The lower failure rate of the RNN-tuned ICCBF suggests that recurrence likely improves robustness to latent parameter variations and accumulated disturbances by exploiting temporal context, similar to the cruise control case. 

\begin{figure}[tbp]
    \centering
\includegraphics[width=\linewidth]{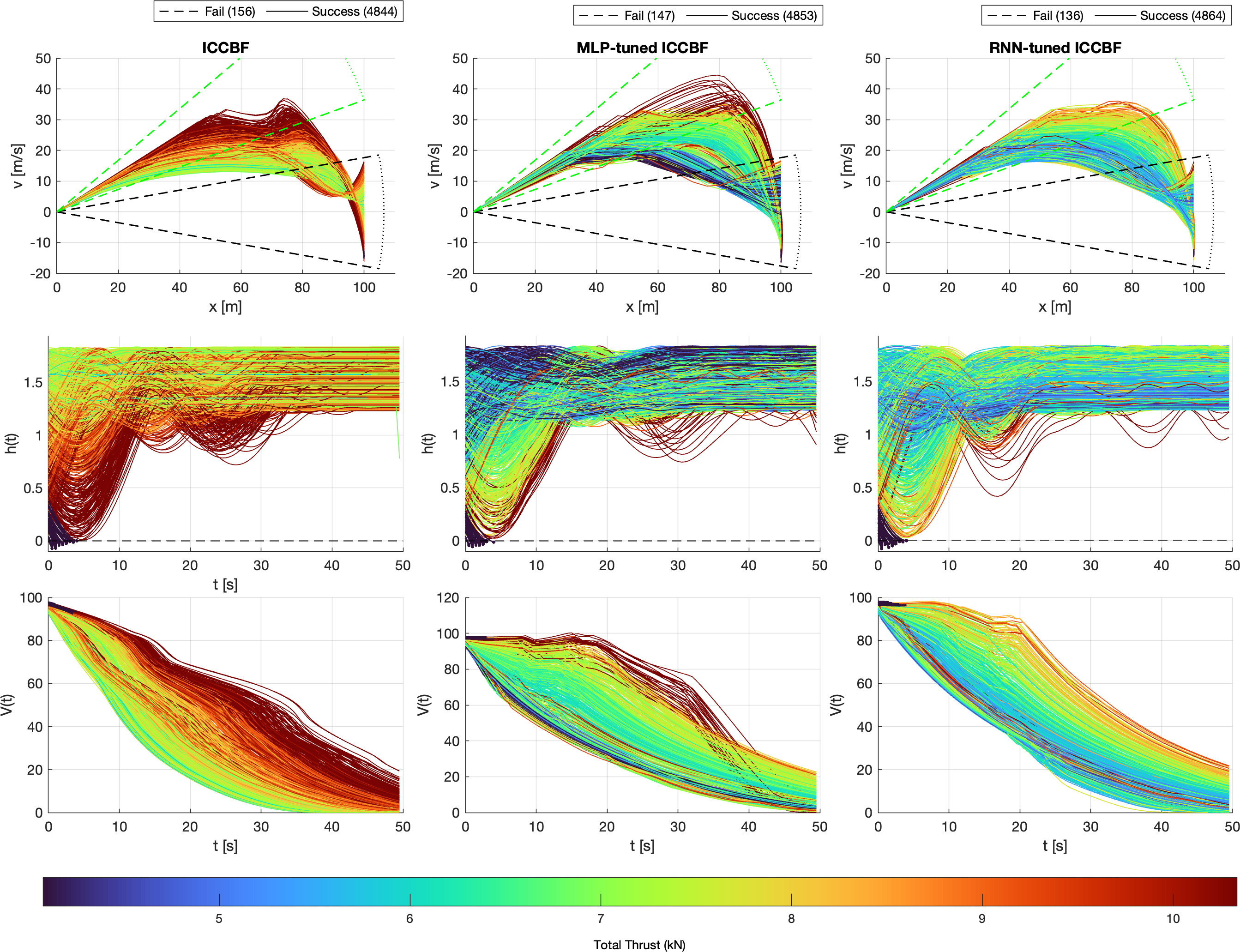}
    \caption{Docking trajectories, CBF and CLF variations over time}
    \label{fig:dock}
\end{figure}

\begin{figure}[tbp]
    \centering
\includegraphics[width=\linewidth]{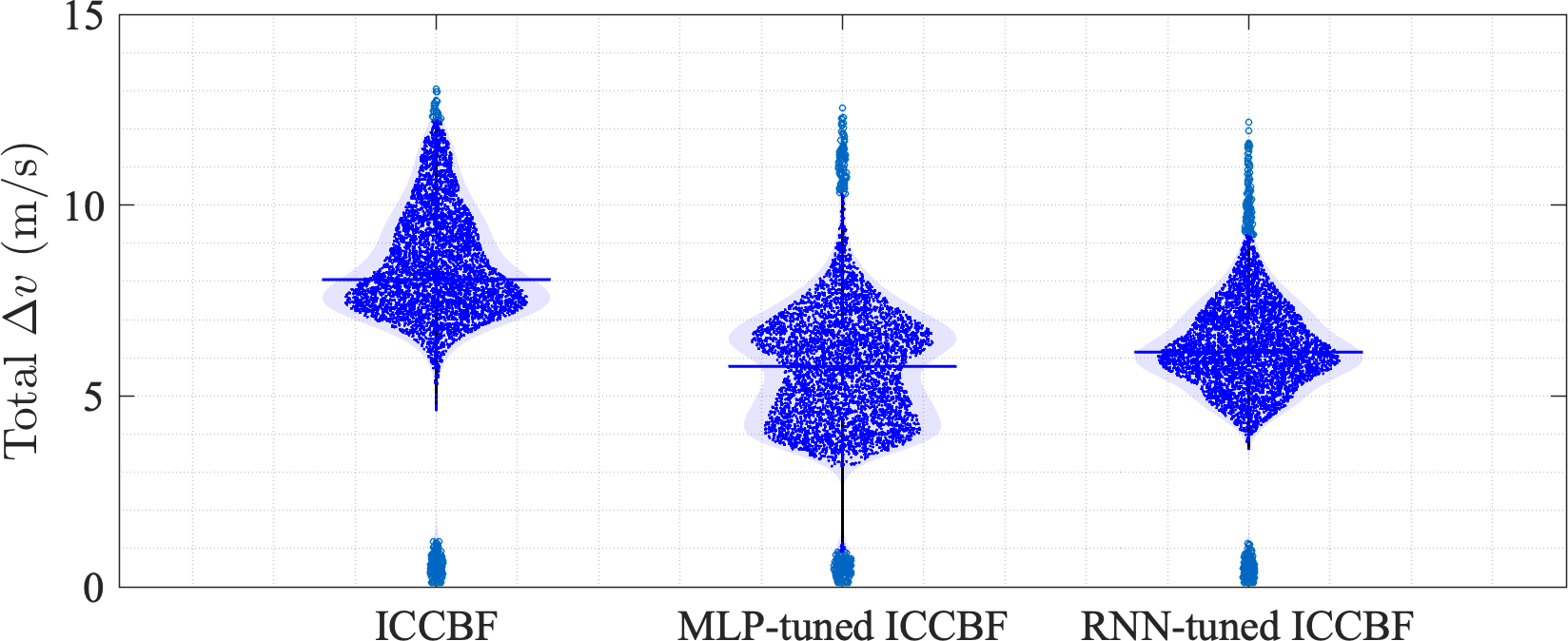}
    \caption{Docking total $\Delta v$ distributions}
    \label{fig:dock2}
\end{figure}


\subsection{3D Spacecraft Inspection}

\begin{figure}[tbp]
    \centering
    \includegraphics[width=0.6\linewidth]{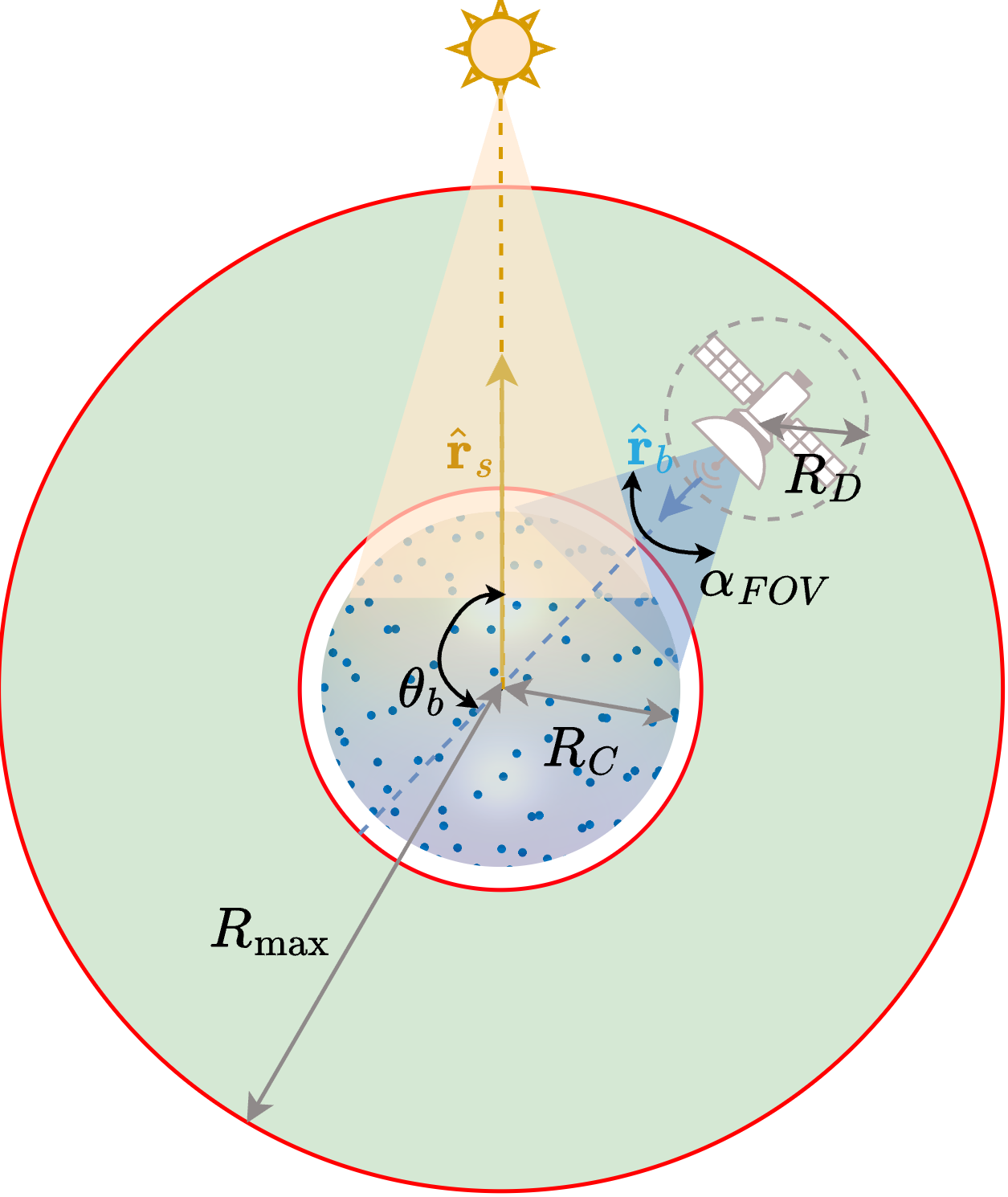}
    \caption{Spacecraft Inspection Problem. $\alpha_{FOV}$ denotes the field of view of the deputy's sensor, while $\theta_b$ is the angle between the sensor boresight and the Sun.}
    \label{fig:inspection3D}
\end{figure}

Lastly, the ICCBF tuning framework is applied to a 3D spacecraft inspection scenario, where a deputy spacecraft must safely maneuver around a spherical chief target while progressively inspecting its surface as shown in Figure \ref{fig:inspection3D}. The chief is modeled as a sphere of radius \(R_C\), and the deputy is assigned an effective collision radius \(R_D\). The chief's surface is discretised into \(N_p\) points, and inspection progress is quantified by the number of points that satisfy admissible viewing and illumination conditions. A reward is provided for each new point inspected.

Note that problem is adapted from \cite{doi:10.2514/1.I011391}, which employs eight first-order CBFs together with velocity and acceleration limits to mitigate degeneracy issues that can arise when safety conditions depend only on position and/or attitude.
In contrast, this work considers only the KOZ, KIZ, and sensor Sun-avoidance constraints, and uses ICCBFs to enforce thrust limits.

\paragraph{Problem Dynamics and Sample Time}
The deputy state is denoted \(\mathbf{x} = [\mathbf{r}^\top,\ \mathbf{v}^\top]^\top \in \mathbb{R}^6\), where \(\mathbf{r}=[x\ y\ z]^\top\) and \(\mathbf{v}=[\dot{x}\ \dot{y}\ \dot{z}]^\top\) are the relative position and velocity expressed in the chief-centred LVLH frame. The relative motion is modelled using the Clohessy--Wiltshire (Hill) equations, which admit a closed-form solution and associated state-transition representation \cite{ClohessyWiltshire1960}. Control is a three-axis thrust input \(\mathbf{u}\in\mathbb{R}^3\) applied under zero-order hold, subject to per-axis saturation \(\|\mathbf{u}\|_\infty \le u_{\max}\).

The sampling period is $T = 10$ s, and the maximum horizon length $N_\text{max}$ is set to 1224 steps to match the conditions in \cite{doi:10.2514/1.I011391}. This results in a maximum mission time of $T_{\mathrm{TOF}} = N_{\max}\,\Delta t = \SI{3.4}{\hour}$, however, the episodes are allowed to terminate early if all points have been inspected under appropriate conditions. 

\paragraph{Initialization}

Episodes are initialized by uniformly sampling the initial range \(\|\mathbf{r}_0\|\in[50,100]~\mathrm{m}\) with random azimuth and elevation, with \(\mathbf{v}_0=\mathbf{0}\) m/s. A sun-angle parameter \(\theta_S\in[0,2\pi]\) is sampled and used to define the sun-direction unit vector \(\hat{\mathbf{r}}^S(\theta_S)\) in the Hill \(x\text{--}y\) plane, going to the Sun from the chief.  The hidden parameters that are varied during each episode include the deputy mass $m$, radius of the deputy $R_D$, radius of the chief $R_C$, $u_{\max}$, $r$, and the KIZ radius $R_{\text{max}}.$ The nominal and relative deviations of these are again given in Table \ref{tab:uncertainty_summary} and shown in Fig.~\ref{fig:inspectionMC}. The standard deviation of thrust uncertainty added is given in Table \ref{tab:noise_models}.

\begin{figure}[tbp]
    \centering   \includegraphics[width=\linewidth]{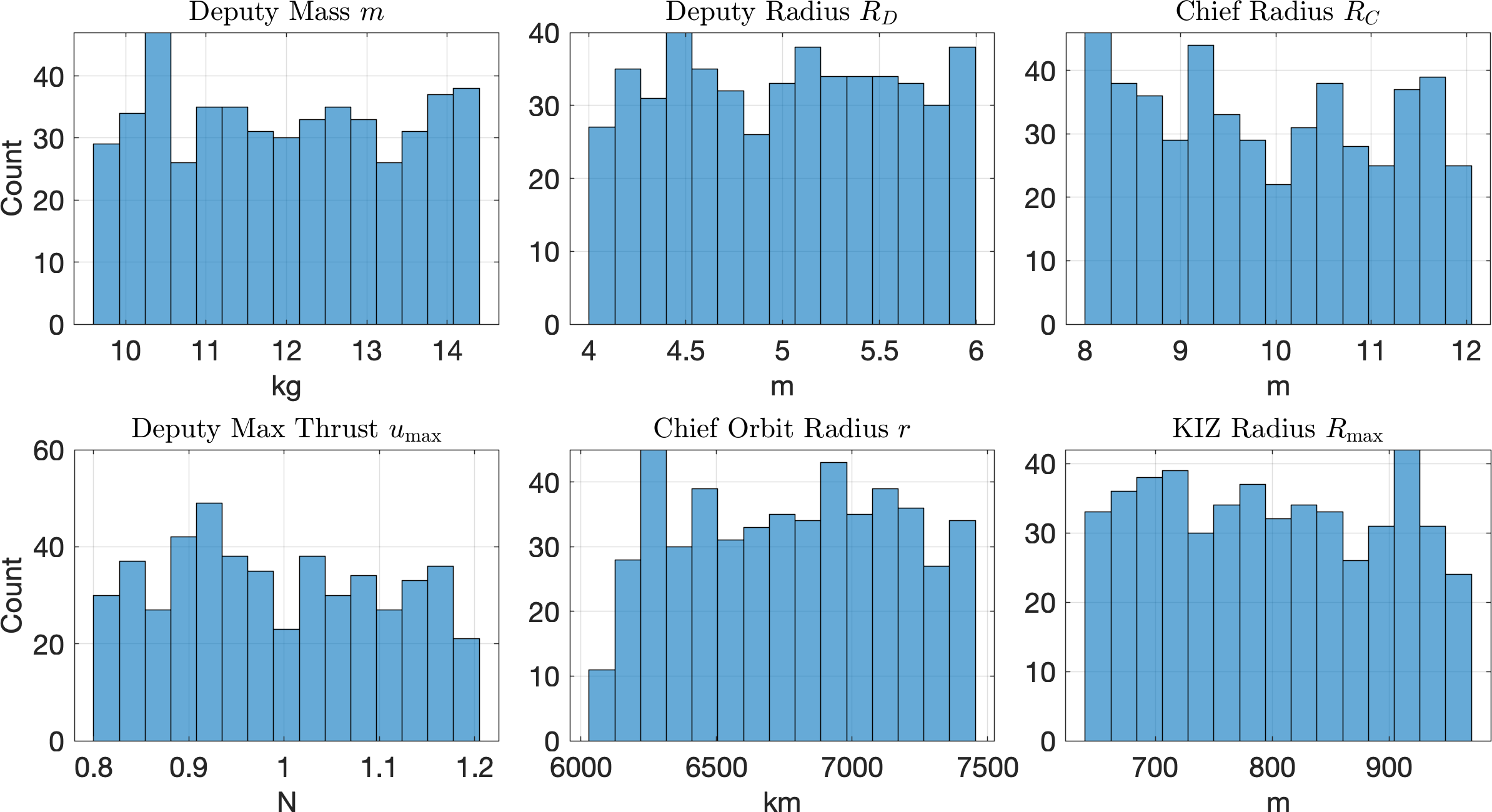}
    \caption{MC parameter variations for 500 inspection samples}
    \label{fig:inspectionMC}
\end{figure}

\paragraph{Observation Model}
As mentioned, a reward is provided based on the number of points that are inspected. A point is considered to be visible if it is in the line of sight of the deputy, inside the field of view angle $\alpha_{FOV}$, the 
point is not in eclipse, and the point is adequately illuminated (not excessively bright or dark). Illumination is evaluated using a Blinn--Phong reflectance model and a shadow gate that requires the surface normal to have a positive dot product with the sun direction. This observation model is discussed in detail in \cite{doi:10.2514/1.I011391}.

\paragraph{ICCBF Safety Filter}

Three safety conditions are encoded into ICCBFs in this work. They are: 
\begin{enumerate}
    \item  Keep-out zone (KOZ) constraint for collision avoidance, where  $ h_1(\mathbf{x}) = \|\mathbf{r}\| - r_{coll} \geq 0$ and $r_{\mathrm{coll}}=R_C+R_D$.
    \item A keep-in zone (KIZ) enforcing $
         h_2(\mathbf{x}) = R_{\max}-\|\mathbf{r}\| \geq 0$.
    \item An angular condition such that the deputy does not align its sensor with the sun $h_3 (\mathbf{x}) = \theta_b - \alpha_{FOV}/2 \geq 0$.
\end{enumerate}
These conditions are enforced through the ICCBF recursion, yielding terminal-stage constraints affine in the control input:
\begin{equation}\label{eq:iccbf_affine}
L_f b_{i,N}(x) + L_g b_{i,N}(x)\,u + \theta_{i,N} b_{i,N}(x) \ge \nu_i(T,x),
\end{equation}
for $i=1,2,3$, where \(\nu_i(T,x)\) are the margins defined in Section~\ref{cmargin}.

Note that unlike with the cruise control and docking cases, the task satisfaction of inspection cannot be written as a CLF in this test case. As such, a learned nominal control policy $\boldsymbol{u}_{RL}$ is required for all methods, including the untuned ICCBF baseline. In the untuned case, ICCBFs act as a supervisory safety filter, modifying the nominal control command to enforce safety. 

At each time step \(k\), the safety filter is posed as the quadratic programme
\begin{align}
\mathbf{u}^\star_k
&=
\arg\min_{\mathbf{u}_k \in \mathbb{R}^m}
\ \|\mathbf{u}_k - \mathbf{u}_{k,\mathrm{rl}}\|_2^2
\label{eq:iccbf_qp} \\
\text{s.t.}\quad
& \text{Eq.~\eqref{eq:iccbf_affine}}\\
& \|\mathbf{u}_k\|_\infty \le u_{\max}.
\label{eq:iccbf_qp_bound}
\end{align}
in addition to $\boldsymbol{\theta}_k = [\theta_{1,0},...,\theta_{1,N},\theta_{2,0},...,\theta_{2,N},\theta_{3,0},...,\theta_{3,N}  ]^T$, the RL actor also provides a nominal thrust RL command \(\mathbf{u}_{\mathrm{k,rl}}\) in this case.

\paragraph{RL State, Reward and Termination Criteria}
The RL state/observation vector is kept the same as in \cite{doi:10.2514/1.I011391}. It includes the deputy relative state with uncertainty $\boldsymbol{x}_k^E$, sun-angle \(\theta_S\), the number of inspected points $n_{insp}$, and a direction vector $\hat{\mathbf{d}}$ pointing toward the largest cluster of remaining uninspected points computed via a \(K\)-means clustering heuristic. In the experiments \(N_p=100\) tiles and \(K=4\) clusters are used.

A reward based on the the number of points inspected is added to the reward given in Eq.~\eqref{CBFreward}, such that 
\begin{align}\label{CBFrewardInsp}
r_k ={}& -w_u \frac{T}{m}\|\mathbf{u}_k^\star\|_2 - w_{\mathrm{fail}}P_{\mathrm{fail}}
       - w_{h_1}\,\mathbb{I}\{h_{1,k+1}<0\}\\ &- w_{h_2}\,\mathbb{I}\{h_{2,k+1}<0\} \nonumber - w_{h_3}\,\mathbb{I}\{h_{3,k+1}<0\}\\
      & + w_{P}\sum \text{Points inspected}.
\end{align}
where $w_u = 0.1$, $ w_{\mathrm{fail}}=   w_{h_1}= w_{h_2} =  w_{h_3}= 1.0$, and $w_P = 0.1$.  Episodes terminate when $P_{\mathrm{fail}} = 1$, which occurs upon QP solver failure or violation of any CBF constraint -or when all inspection points have been successfully inspected.

\paragraph*{Outcomes}
The inspection results are provided in Figures~\ref{fig:insp}, \ref{fig:insp2}, and Table~\ref{tab:performance_summary}, and clearly show the advantage of incorporating recurrence when tuning CBFs. The RNN-tuned ICCBF exhibits the strongest overall performance, achieving near-complete inspection coverage while consuming the least fuel. This behavior suggests that the recurrent policy can exploit temporal context and infer latent structure in the inspection geometry that is not directly observable at a single time step.

The baseline untuned ICCBF employs fixed barrier weights ($\boldsymbol{\theta} = 0.05\,\mathbf{1}_9
$); however, the applied control input \(u_{k,\mathrm{RL}}\) is still learned in this case. While this approach yields high inspection completion in the majority of cases, it does so at the cost of the largest total \(\Delta v\), indicating that robustness and observability are achieved through conservative control effort. In contrast, the MLP-tuned ICCBF more aggressively minimizes \(\Delta v\), but this reduction comes at the expense of inspection performance. A significant fraction of episodes terminate with incomplete coverage, indicating that the feedforward policy tends to prioritize fuel efficiency over task completion under the chosen reward structure. The RNN-tuned ICCBF provides the most favorable trade-off. The inspection-score distribution is concentrated at \(100\%\), with all quartiles at full coverage, implying that at least \(75\%\) of the episodes achieve complete inspection. Simultaneously, the total fuel consumption remains low, with a \(P_{99}\) of only $\SI{15.27}{\meter\per\second}$, highlighting strong tail performance as well as consistency across trials. Both the untuned case and the RNN-tuned case retains safety across all cases tested.

Finally, it is noted that these results are inherently influenced by the relative weighting between inspection progress and \(\Delta v\) minimization in the reward function. The trends reported here correspond to the constant-weight configuration used in the inspection experiments; alternative weighting strategies would shift the balance between task completion and fuel optimality.

\begin{figure*}[tbp]
    \centering
\includegraphics[width=\linewidth]{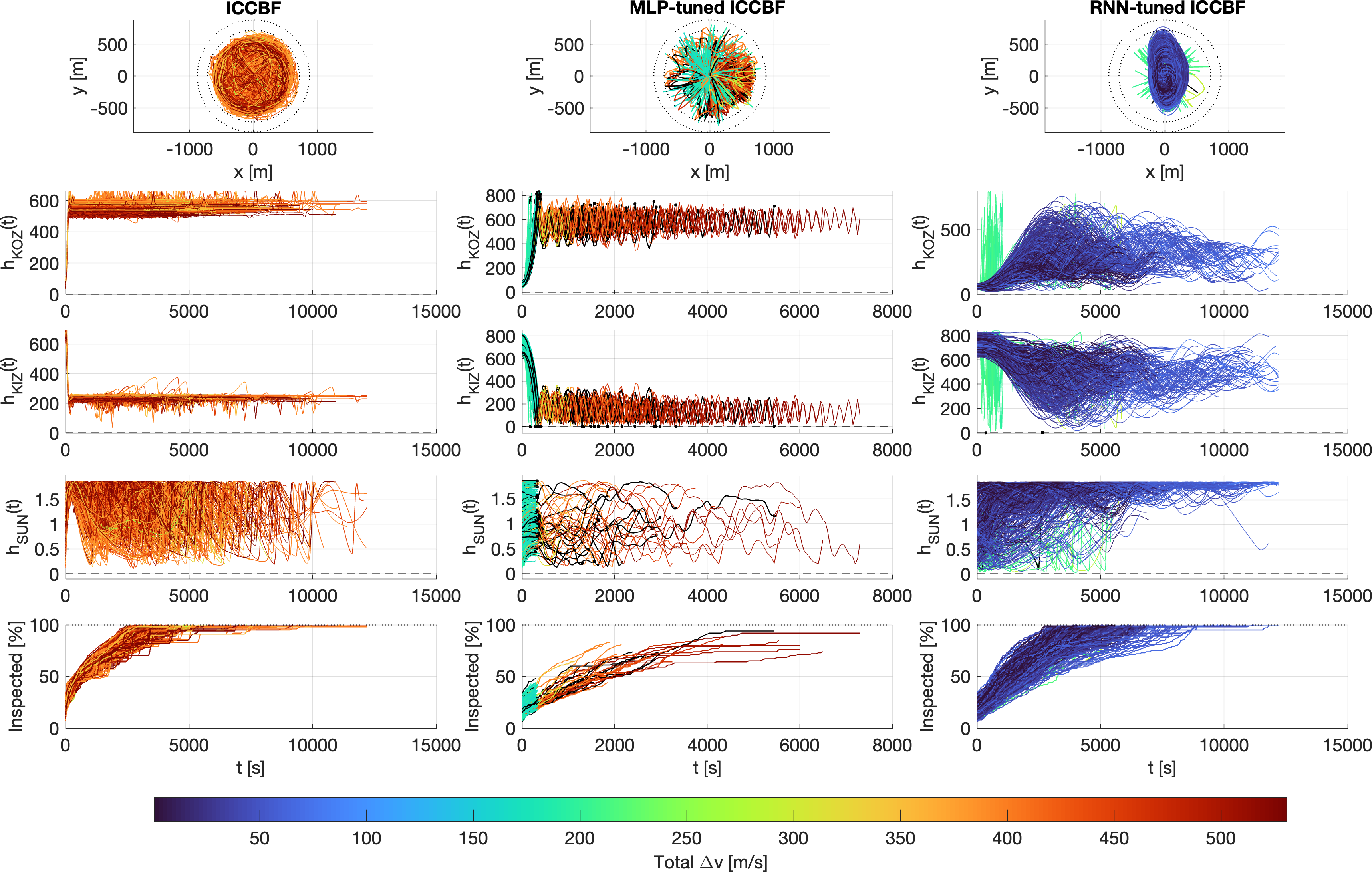}
    \caption{Inspection trajectories, CBF and inspection score variations over time}
    \label{fig:insp}
\end{figure*}

\begin{figure}[tbp]
    \centering
\includegraphics[width=\linewidth]{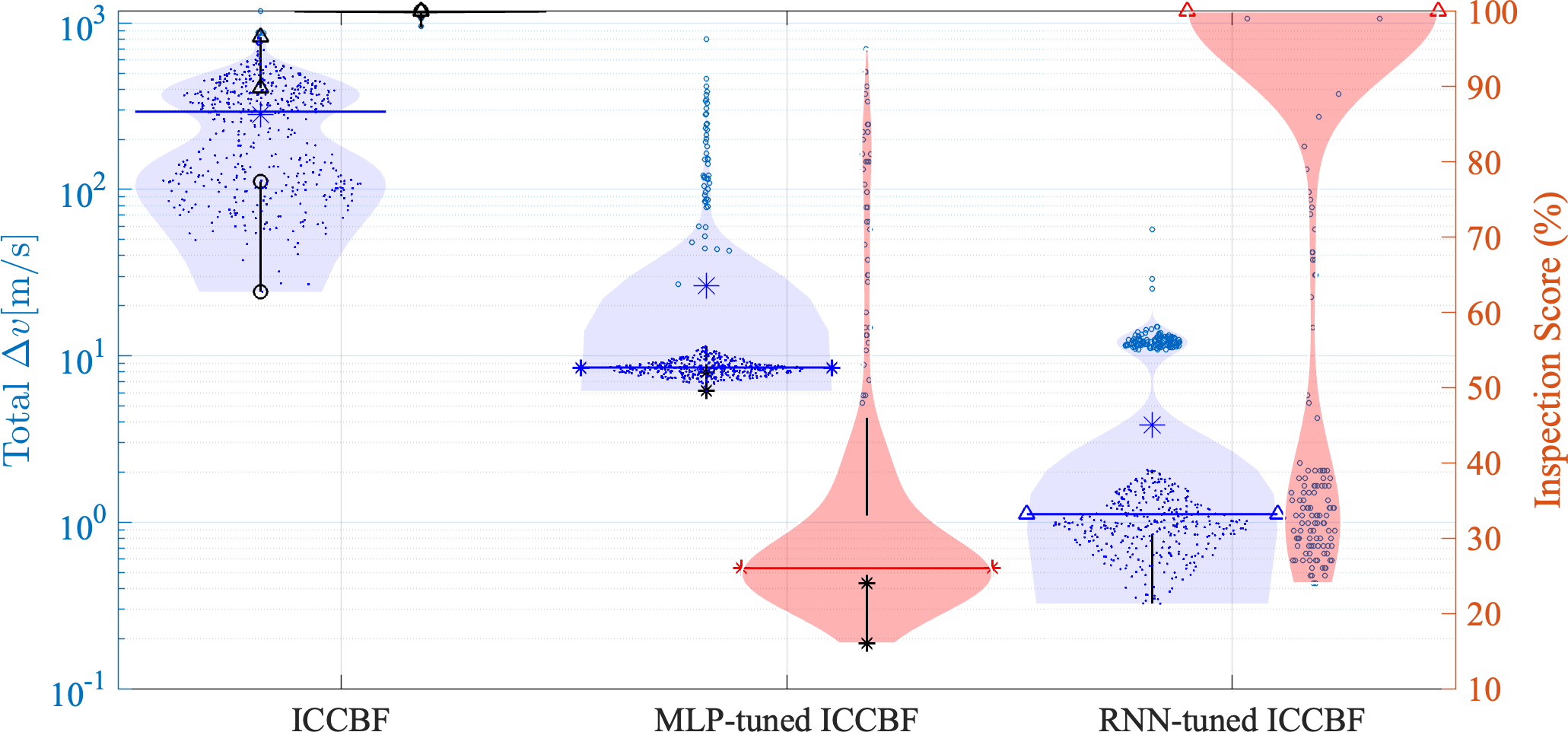}
    \caption{Inspection total thrust and score distributions}
    \label{fig:insp2}
\end{figure}

\section{CONCLUSION}\label{sec:conclusion}

This work develops a meta-RL framework to tune ICCBFs for non-myopic behavior and enhanced robustness to state uncertainty, thrust limitations, and hidden-parameter variations. Robustness to hidden parameters is particularly critical in PRO, where properties of the target object are often known with some uncertainty. The proposed approach learns the hierarchy of local class-$\mathcal{K}$ functions used to derive continuous-time ICCBFs, enabling the controller to balance safety, fuel optimality, and feasibility while retaining empirical safety guarantees. The learned tuning is implemented on a time-sampled dynamical system using a control margin efficiently computed via DA, preserving on-board computational tractability.

Across representative cruise control, docking, and inspection scenarios, the results demonstrate that recurrent architectures such as LSTMs are particularly effective for tuning ICCBFs. The RNN-tuned ICCBF consistently achieves lower fuel consumption while maintaining a larger set of feasible and successful trajectories, especially in cases with more hidden parameters. Particularly in the inspection problem, the MLP-based approach frequently trades inspection completeness for reduced $\Delta v$, whereas the RNN successfully learns non-greedy inspection trajectories that preserve constraint satisfaction, lower fuel consumption, and improve inspection score. These findings highlight the importance of temporal memory when learning safety-critical control laws in structured but partially observed environments.

\bibliographystyle{abbrv}
\bibliography{refs}

\end{document}